\documentclass[12pt,a4paper]{article}

\usepackage[english]{babel}

\usepackage{graphicx}
\usepackage{amsmath,amsfonts,amssymb}
\usepackage{amsbsy}
\usepackage{mathrsfs} 
\usepackage{varioref,hyperref}
\usepackage[shortlabels]{enumitem}
\usepackage{bm}
\usepackage{float}
\usepackage{algorithm}
\usepackage{algorithmicx}
\usepackage{algpseudocode}
\usepackage{listings}
\lstdefinestyle{cpp}{
	language=c++,
	breaklines=true,
	basicstyle={\small\ttfamily},
	showstringspaces=false,
	showtabs=false,
	keepspaces=false,
	columns=fullflexible,
	tabsize=1
}
\usepackage{placeins}
\usepackage[dvipsnames]{xcolor}

\usepackage[font=small]{caption}
\usepackage{epsfig}
\usepackage{array}
\usepackage{verbatim}
\usepackage{graphicx}
\usepackage{color}
\usepackage{tabularx}
\usepackage{bm}
\usepackage{booktabs}
\usepackage{subfig}
\usepackage{stmaryrd}
\usepackage{textcomp}

\usepackage{array}

\usepackage{multirow}
\usepackage{tabularx}
\usepackage{url}

\begin{document}

\title{Portable, Massively Parallel Implementation of a Material Point Method for Compressible Flows}
\author{Paolo Joseph Baioni$^{(1)}$, Tommaso Benacchio$^{(2)}$, \\Luigi Capone$^{(3)}$, Carlo de Falco$^{(1)}$}

\maketitle

\begin{center}

{\small
\vskip 0.5cm
$^{(1)}$ MOX -- Modelling and Scientific Computing \\
Dipartimento di Matematica, Politecnico di Milano \\
Piazza Leonardo da Vinci 32, 20133 Milano, Italy\\
{\tt \{paolojoseph.baioni,carlo.defalco\}@polimi.it} }\\

{\small
\vskip 0.5cm
$^{(2)}$ Danish Meteorological Institute \\
Sankt Kjelds Plads 11 \\
2100 Copenhagen, Denmark\\
{\tt tbo@dmi.dk}
}\\

{\small
\vskip 0.5cm
$^{(3)}$  Leonardo Labs, Leonardo S.p.A.\\
Torre Fiumara, Via Raffaele Pieragostini, 80, 16149 Genova, Italy \\
{\tt luigi.capone@leonardo.com}
}
\end{center}

\date{}

\vskip 0.5cm


\vspace*{0.5cm}


\newpage
\vfill
{\noindent\footnotesize This preprint is a preliminary version of a paper published in Computational Particle Mechanics, doi: \texttt{https://doi.org/10.1007/s40571-024-00864-2}. URL of the published version: \texttt{https://link.springer.com/article/10.1007/s40571-024-00864-2}.}

\pagebreak

\abstract{The recent evolution of software and hardware technologies is leading to a renewed computational interest in Particle-In-Cell (PIC) methods such as the Material Point Method (MPM). Indeed, provided some critical aspects are properly handled, PIC methods can be cast in formulations suitable for the requirements of data locality and fine-grained parallelism of modern hardware accelerators such as Graphics Processing Units (GPUs). Such a rapid and continuous technological development increases also the importance of generic and portable implementations. While the capabilities of MPM on a wide range continuum mechanics problem have been already well assessed, 
        the use of the method in compressible fluid dynamics has received less attention. 
		In this paper we present a portable, highly parallel, GPU based MPM solver for compressible gas dynamics. The implementation aims to reach a good compromise between portability and efficiency in order to provide a first assessment of the potential of this approach in solving strongly compressible gas flow problems, also taking into account solid obstacles. The numerical model considered constitutes a first step towards the development of a monolithic MPM solver for Fluid-Structure Interaction (FSI) problems at all Mach numbers up to the supersonic regime.}

\vfill
\section{Introduction and Motivation} \label{intro}
	\noindent Hardware and software technology evolution has led to a renewed interest in particle-based numerical techniques for differential problems, such as the Material Point Method (MPM,~\cite{SulskyChenSchreyer1994,MPMBook}), that was developed stemming from the \textit{Particle In Cell} method \textit{Fluid Implicit Particle Method}~\cite{Harlow,FLIP} to treat history-dependent materials, thanks to their suitability to modern hardware accelerators requirements~\cite{Hariri,RealTime}. Indeed, parallel computing is a key capability for effective advancement of industrial-grade, large-scale computations, particularly in the fields of high-speed, compressible computational fluid dynamics and Fluid-Structure Interaction (FSI,~\cite{FutureDirections}) simulations. On the other hand, rapid hardware and software evolution makes researchers lean towards generic and portable implementations, which enable developers to swiftly adapt to sudden technology innovations.\\
	In view of exponentially increasing High Performance Computing (HPC) resources and the method's accurate handling of a wide range of phenomena, MPM is becoming increasingly relevant for compressible CFD and FSI simulations. In MPM, continuum media are discretised by means of Lagrangian particles, the \textit{material points}, each one storing every state variable evaluated at its position, enabling the description of a wide range of materials~\cite{Goktekin,Stomakhin,CJangthesis} as well as large deformations~\cite{LargeDeform1,LargeDeform2,TLMPM} and fracture phenomena~\cite{fratture}. Notwithstanding its Lagrangian character, MPM also uses a background Eulerian grid, that is often chosen to be a tensor-product Cartesian grid for sake of efficiency, to compute differential quantities and solve the motion equation, thus mediating the particle-particle interactions, and taking advantage of both Eulerian and Lagrangian approaches. These features make MPM an interesting candidate for the development of a highly parallel, Graphics Processing Units (GPUs)-based FSI solver. 
	\noindent Indeed, in order to reach a high degree of efficiency, GPU-based solvers require a large degree of data locality and fine-level parallelism, and MPM implementations can naturally achieve these, since: 
	\begin{itemize}
		\item every GPU thread can manage up to very few grid cells or particles;
		\item the link between cells and particles can be easily built thanks to the structured character of the grid;
		\item accuracy in the discretisation of continuum bodies is obtained by means of the Lagrangian particles, so that the grid does not have to match the solid boundary.
	\end{itemize}
	Thus, there is no need of computationally expensive, unstructured, moving or deforming meshes and a structured Cartesian grid can be used instead, simplifying also techniques such as Adaptive Mesh Refinement and Domain Decomposition in a parallel context~\cite{adapt1,MultiGPU}.\\
	Since the first introduction of MPM, various enhancements of the method were proposed, solving some of its drawbacks such as grid-crossing instabilities~\cite{Tielen} and volumetric locking~\cite{Coombs}, and improving its accuracy and conservation properties. To address instabilities which arise when particles cross the grid, variants of the method were proposed, such as GIMP~\cite{GIMP}, CPDI~\cite{CPDI}, iMPM~\cite{iMPM} and IGA-MPM~\cite{IGAMPM}, and numerical analysis studies on the use of B-Splines as basis functions in MPM were conducted~\cite{Steffen}. An affine and a polynomial mapping were later added to the B-Splines MPM~\cite{APIC,PolyPIC} resulting in an angular momentum-preserving models. Other studies explored least squares techniques~\cite{Wobbes} enabling the recovery of the affine/polynomial splines MPM~\cite{MLSMPM}, the effect of spatial and temporal discretisation errors, and the application of symplectic integrators~\cite{Berzins}. In light of these developments, MPM represents an effective method for the simulation of a wide range of materials and phenomena.\\
	However, even though MPM was actually initially developed for fluid simulations, there are only a few recent studies addressing compressible fluid dynamics using MPM. \\
	Existing studies on MPM for compressible flows focused on modelling, as the first work addressing compressible fluids and thin membranes~\cite{York}, on numerical convergence and error analysis~\cite{Tran}, and on expanding the range of applications for the method, as the more recent paper~\cite{ChenTD}, which also takes into account thermomechanics effects; finally, in~\cite{Schroeder2024} the use of PolyPIC transfers~\cite{PolyPIC} to construct a second order accurate discretisation of the Navier-Stokes equations within a PIC-MPM context is proposed.
	Nonetheless, none of these studies addresses high speed gas dynamics. In addition, the existing studies, some of which (e.g.~\cite{York}) predate work on numerical linear algebra on GPUs (e.g.~\cite{gpgpu2003,gpgpu2005}), do not deal with parallelisation on modern hardware.\\
	In fact, in recent years MPM has gained popularity in the computer graphics community (e.g.~\cite{StomakhinSnow}), leading to relevant results on modern hardware. In particular, Gao et al.~\cite{GPU} designed an MPM implementation optimised for GPUs, adapting the efficient data structure for sparse data on CPU of Setaluri et al. (\cite{spgrid}), and Wang et al.~\cite{MultiGPU} extended the work to a multi-GPU shared memory, MPI-free programming model.
    Finally, Buckland et al.~\cite{AdV2024} recently considered the problem of portability of a parallel MPM implementation in the computational mechanics realm.\\
	However, these works do not treat compressible fluid dynamics problems, and the choice of the material to simulate has additional consequences on the computational efficiency. \\
	For example, the GPU-optimised SPGrid data structure of~\cite{GPU} is designed for a computational domain with a relevant portion of empty cells, as it employs MPM to simulate solids, elasto-plastic materials, and sand. Conversely, gases fill the computational domain, requiring a different data structure. The efficiency of MPM on modern, parallel hardware in the modelling of compressible gas dynamics has yet to be proved.
	Indeed, existing studies addressing Fluid-Structure Interaction in a compressible, high speed gas dynamics context typically rely on a segregated approach, coupling different solvers for different materials. This approach, besides the known pros and cons with respect to monolithic solvers, also introduces challenges in the performance portability of the implementation. \\
	A very good example of this approach is given by~\cite{gas}, where MPM has been used only for the solids, while the compressible fluid dynamic problem has been solved through a second order WENO scheme, and the code is run on CPUs. Indeed, high-order WENO and ENO schemes guarantee numerically excellent results in gas dynamics but, for these schemes to work properly, the stencil should be able to grow in any direction into neighbouring mesh elements. From a computational point of view, this works very well on structured grids. However, designing proper structured grids for Eulerian methods in FSI is significantly time consuming. On the other hand, if unstructured grids are chosen, growing stencils leads to a large increase of halo transfers in communication among parallel processors, thus requiring a large consumption of memory bandwidth and generating a large amount of indirection, degrading the suitability of the methods for modern parallel hardware.\\
	In this realm, MPM presents very interesting features as complete FSI solver, indeed: 
	\begin{itemize}
		\item it benefits from the choice of employing structured, Cartesian tensor product grids, since continua discretisation is delegated to Lagrangian material particles, allowing the modelling of large deformations without the need of complex and costly meshes;
		\item it does not require coupling among solvers, as it is suitable for structural mechanics problems as well. 
	\end{itemize}
	Nonetheless, to the best of the authors' knowledge, no studies exist that are dedicated to the GPU based, High Performance Computing aspects of MPM implementations for strongly compressible fluid dynamics problems. Our aim is to explore these aspects in depth, considering also the performance portability features of the algorithm and its implementation.\\
	With this work we aim to address only the fluid portion of a future FSI solver, i.e., the portion which is usually left to Eulerian CFD solvers. We focus on algorithm and data structures suitable for implementation on various parallel architectures, with a particular focus on HPC GPUs, and we address the problem of reducing limitations in treating obstacles that derives from the choice of a structured Cartesian grid, by means of a signed distance function approach. \\
	The current study focuses on linear MPM to better study its scalability features, leaving the update to a higher order MPM for future work. However, we anticipate a discussion on how higher regularity base functions could fit our approach and how the parallel efficiency would be influenced by the extension of base function support on more elements. From the modelling point of view, to give a first assessment of MPM in simulating efficiently compressible gases on GPUs, we treat the gas as inviscid fluid, neglecting the effects of turbulence, and adopt an artificial viscosity based shock-capturing technique.\\ 
	In previous work~\cite{particles}, we ported an existing MPM algorithm~\cite{York,YorkThesis,Sandia} to CUDA-C~\cite{CUDAC}, adapting it to GPU architecture and obtaining promising speed-ups with respect to a C++ CPU version. The CUDA-C programming model was chosen because of its known performance capabilities on NVIDIA A100 GPUs~\cite{A100,GPUcfr}. However, the programming model had shortcomings in its low-level character and specificity to the underlying architecture: while it can be made to work on, e.g., AMD GPUs via HIP with minimal changes, it cannot be ported to different architectures such as multicore CPUs without substantial code modifications.\\
	One of NVIDIA's solutions for performance portability involves its own C++ compiler nvc++~\cite{nomoreporting,NvidiaNvc++}, using which, one rewrites algorithm steps relying on C++ Standard Template Library (STL), specifying a parallel execution policy. To give an example, in this framework it is  possible to parallelise the following for loop:
	\begin{lstlisting}[style=cpp, firstnumber=1, numbers=left, numberstyle=\tiny,showtabs=false]
		for (index_t i = 0; i < n; ++i)
		{do some operations} 
	\end{lstlisting}
	into a call to a {\ttfamily for\_each} with a proper lambda or functor, e.g.:
	\begin{lstlisting}[style=cpp, firstnumber=1, numbers=left, numberstyle=\tiny,showtabs=false]
		std::for_each (std::execution::par, first_counting_iterator, last_counting_iterator, [](index_t i){do some operations});
	\end{lstlisting}
	Although the approach does not prevent the developer from introducing data-races or deadlocks, code can be written only once, and then, depending on options specified at compile time, run in parallel on multi-core CPUs or GPUs.\\
	However, nvc++ high-level features come at a price, as the developer loses the control on memory handling and, in particular, on the moment when Host-to-Device and Device-to-Host memory copies are performed. The code runs on host, any invocation to an algorithm with a parallel execution policy expresses the developer's preference to run that algorithm in parallel. If compiled for GPUs, memory is copied to the device, computations are performed and then data is copied back to host. In physics and in engineering dynamics simulations, where each time step depends on the previous one, this means that the maximum reachable level of parallelisation corresponds to a single call to a parallel algorithm inside a traditional {\ttfamily for} loop for time advancing. Since memory fetches can be significantly more costly than computations, this is sub-optimal.\\ 
	The present contribution expands the scope of~\cite{particles} and overcomes the performance issues with nvc++ by designing a portable MPM code for transonic and supersonic compressible gas dynamics based on Thrust~\cite{Thrust}, a template library for CUDA and HIP based on the C++ Standard Template Library (STL). 
	In addition to nvc++ features, Thrust introduces two types of STL like vectors, one for the host (typically, the CPU) and one for the device (typically, the GPU); this allows the user to specify when memory should be copied via explicit calls. For example, it is possible to perform a single Host to Device copy when reading the initial data file, and one Device To Host copy whenever saving partial results on output data files, strongly reducing execution time related to memory fetches. \\
	Moreover, as additional portability-enhancing features included in the implementation:
	\begin{itemize}
		\item Thrust syntax is mostly compatible with STL syntax, adding just a few {\ttfamily define} preprocessor macros or templates. Thanks to this, almost the same source code can be compiled for parallel execution on GPUs with NVIDIA CUDA compiler nvcc, on AMD GPUs with the AMD compiler for ROCm/rocThrust, and on CPUs with g++ and clang++ compilers (when it is linked with a parallel library such as TBB, Intel's Threading Building Blocks).
		\item Thrust itself allows for modifications to the default device and host system, so that it can be run both serially and in parallel, with TBB or OpenMP on CPUs for the host, and also on the GPU via CUDA-C++ for the device only. The code can be compiled and executed on a CPU-only machine as well, provided  {\ttfamily THRUST\_DEVICE\_SYSTEM} is defined to be\\ 
  {\ttfamily THRUST\_DEVICE\_SYSTEM\_\{CPP,OMP,TBB\}} at compilation stage.
	\end{itemize}
	\noindent Specific development choices are described that enable optimisation on modern GPU-based HPC architectures and portability to different hardware, as well as generality of the approach for the purposes of its adaptability also towards different software. The accuracy and performance of the implementation are evaluated on a number of compressible flow benchmarks and the results are compared with available approaches in the literature.\\
	The paper is structured as follows. Section~\ref{sec:num} contains the analytical and numerical formulation of the method. Section~\ref{sec:alg_and_impl} describes the chosen algorithm and the current and implementation choices. Section~\ref{sec:res} illustrates the results obtained on benchmark test cases, addressing numerical modelling aspects as well as performance portability, and the final section~\ref{sec:concl} contains a discussion of the results, draws the conclusions of the performed work, and prospects future developments.
	
	\section{Model Formulation}\label{sec:num}

 \noindent{In this section we present, in~\eqref{eq:fullmodel} of subsection~\ref{ssec:modelequations}, the model equations that are actually solved by our code. In order to clearly point out the simplifications that we adopted and the physical limitations they correspond to, we briefly outline the derivation of such model from the more general form stated in~\eqref{eq:euler_0}.
 
 \noindent As previously mentioned, MPM takes advantages of both the Eulerian and the Lagrangian perspective; specifically, the Lagrangian particles discretisation approach affords flexibility in modelling various kind of phenomena and materials, while the Eulerian grid discretisation is beneficial for scalability as the grid allows the mediation of particle-particle interactions.

 \noindent As the Eulerian approach used for the integration of the moment conservation equation is based on the weak form of such equation, the weak formulation is presented at the end of subsection~\ref{ssec:modelequations}, before proceeding to the description of the numerical algorithm in subsection~\ref{ssec:numericalmethod}

 \subsection{Model Equations and Weak Formulation}
\label{ssec:modelequations}
 
    \noindent Denoting by $\dfrac{d \cdot }{d t}$ the \textit{material 
    derivative} operator, the flow of a compressible fluid characterised  by its mass density $\rho$, velocity $\mathbf{v}$ and 
    internal energy $e$ and contained in a domain $\Omega$ is
    governed by the following set of conservation laws 
   ~\cite{MaierLike,YorkThesis,AdVreview}:
		\begin{equation}
			\label{eq:euler_0}
			\begin{cases}
				\dfrac{d \rho}{d t} + \rho \nabla \cdot \mathbf v = 0\\[2mm]
				\rho\dfrac{d \mathbf v}{d t} = \nabla \cdot \bm\sigma  + \rho\mathbf b\\[2mm]
				\rho\dfrac{de}{dt} = \bm \sigma : \bm {\dot\varepsilon} - \nabla \cdot \mathbf h + \rho s.
			\end{cases}
		\end{equation}
        In the equations above, $\bm \sigma$, $\rho \mathbf b$,
        $\bm{\dot\varepsilon}$, $\mathbf h$ and $s$ represent  the stress tensor, the external force density, the strain rate tensor, the heat flux and the heat source, respectively; suitable constitutive relations for such quantities must be provided in order for the model to be solvable, our choices for such relations are described below.
        For the stress tensor $\bm{\sigma}$ we assume the 
        following Ansatz
        \begin{equation}\label{eq:sigma=-pI}
            \sigma_{\alpha\beta}= - (p + q)\ \delta_{\alpha\beta}
	\end{equation}
        \noindent where $\delta_{\alpha\beta}$ is the Kronecker delta.\\
        In~\eqref{eq:sigma=-pI}, $p$ denotes the thermodynamic 
        pressure which we assume to be related to the state 
        variables via the state equation for a perfect gas
        \begin{equation}
				p = (\gamma -1)\rho e
		\end{equation}
        where $\gamma$ is the ratio between specific heat at constant pressure and constant volume which, for 
        a bi-atomic perfect gas, is given by $\gamma=c_p/c_v=\tfrac{7/2}{5/2} = 1.4$.
        While viscous stresses are neglected in the expression
        of~\eqref{eq:sigma=-pI} (indeed we focus on inviscid 
        fluid flow only in the present work), a small
        artificial viscosity is commonly adopted in numerical 
        simulations to improve stability. In this paper this
        is accomplished, as suggested in~\cite{MPMBook}, 
        via the additional pressure term $q$ in~\eqref{eq:sigma=-pI} 
        which consists of a combination of von Neumann-Richtmyer~\cite{mu_art_1} and 
        Landshoff~\cite{mu_art2} artificial viscosities and is meant to damp out
        spurious oscillations near solution discontinuities. 
        The artificial viscosity correction is only activated in compression regions, 
        identified as areas where the divergence of $\mathbf{v}$ is less than zero, 
        and its expression reads
	    \begin{equation}\label{eq:muart}
		q = 
		\begin{cases}
			c_0 \left(\rho\,h\,\mathrm{div}(\mathbf v)\right)^2 - c_1\,h\,\sqrt{\gamma\tfrac{p}{\rho}}\,\mathrm{div}(\mathbf v) \,\, &\mathrm{div}(\mathbf v)<0\\
			0 \quad &\mathrm{div}(\mathbf v)\geq 0
		\end{cases}
	     \end{equation}
	   where $c_0,c_1$ are dimensionless constants and $h$ denotes 
      the characteristic length of a grid cell. 
      In section~\ref{sec:res} we assess the effect of the different terms
      comprising the correction $q$ on the solution of a problem displaying a shock wave, a contact discontinuity and a rarefaction wave.
    Finally, we consider an adiabatic, non reactive flow regime and, therefore, 
    neglect heat fluxes and sources, as well as external body forces.\\

	\noindent In conclusion, the system of equations we focus on in the present study can 
   be summarised as
    \begin{subequations}\label{eq:fullmodel}
	\begin{equation}
		\label{eq:euler}
		\begin{cases}
			\dfrac{d \rho}{d t} + \rho \nabla \cdot \mathbf v = 0\\[2mm]
			\rho\dfrac{d \mathbf v}{d t} = - \nabla (p+q) \\[2mm]
			\rho\dfrac{d e}{d t} = -(p+q) \nabla \cdot \mathbf v\\[2mm]
 		\end{cases}
	\end{equation} 
    and
   \begin{equation}\small
		\label{eq:pressures}
		\begin{cases}
			p = (\gamma -1)\rho e\\[2mm]
            q = c_0 \left(\rho\,h\,\mathrm{div}(\mathbf v)\right)^2 - c_1\,h\,\sqrt{\gamma\tfrac{p}{\rho}}\,\mathrm{div}(\mathbf v),  &  \mathrm{div}(\mathbf v)<0\\
	        q = 0,  &  \mathrm{div}(\mathbf v)\geq 0.
    \end{cases}
	\end{equation} 
   \end{subequations}
    Of course~\eqref{eq:fullmodel} must be supplemented by suitable initial and boundary 
    conditions, which will be detailed separately for each of the individual flow configurations 
    considered in the following.\\
	This dynamically comprehensive yet relatively simple equation set is chosen as it enables testing the performance portability of the MPM implementation on compressible gas dynamics benchmarks~\cite{Toro,Sod} and comparison with existing approaches~\cite{York,ChenTD} (see eqs. (2.55)-(2.61) in~\cite{MPMBook} for a comparison). \\
	As shown in the next section, the momentum equation is solved on the grid, while mass conservation, internal energy conservation and state equations are solved on particles (see step 10 in~\ref{algo1}.).
	The link between the Lagrangian particles and the Eulerian mesh is given by mappings which can be derived from the weak formulation of the motion equation~\cite{SulskyChenSchreyer1994,MPMBook}, that is obtained as follows.\\
	Let $\Omega \in \mathbb{R}^2$ be a domain, and $\mathbf {\dot v}:\Omega \rightarrow \mathbb{R}^2$ denote the total, material, derivative of the velocity. Then, momentum conservation reads
	\begin{equation}\label{eq:momentum}
		\rho \mathbf {\dot v} = - \nabla (p+q)
	\end{equation}
	Take a test function $\mathbf w \in V = [H_0^1(\Omega)]^2$, multiply~\eqref{eq:momentum} by $\mathbf w$ and integrate over the domain $\Omega$
	\begin{equation}
		\int_\Omega \rho \mathbf{\dot v} \cdot \mathbf w d\Omega = -\int_\Omega  \nabla (p+q) \cdot \mathbf w d\Omega
	\end{equation}
	Integrating by parts the second term
	\begin{equation}\label{eq:weak_form}
		\int_\Omega \rho \mathbf{\dot v} \cdot \mathbf w d\Omega = \int_\Omega (p+q) \nabla \cdot \mathbf w d\Omega
	\end{equation}
	where boundary integrals are neglected because of the requirement that $\mathbf w \in V = [H_0^1(\Omega)]^2$.\\
	The continuity, the energy conservation and the state equations will be directly discretised and used to update particle properties in the algorithm (see section~\ref{sec:alg_and_impl} and algorithm~\ref{algo1}), while the discretisation of the weak form of the momentum equation, related to the link between the Eulerian grid and the Lagrangian particles, is treated in more depth in the next section.\\

\subsection{Numerical Method}
\label{ssec:numericalmethod}

	In this subsection we derive the formulae which serve as Particle-to-Grid mappings from the discretisation of the weak form of the motion equation~\eqref{eq:weak_form}.
	Then, building upon the derived expression, we define four mapping operations in a general way, such that they can be applied to different quantities to formulate the algorithm.\\
	This step is indeed fundamental, since by expressing the whole algorithm in terms of four main kernels we are then able to perform better computational analysis, measures and optimisation.\\
	In the following, we proceed in the afore-mentioned Galerkin way, as discussed also in~\cite{Sandia} and done by~\cite{SteffenThesis,Steffen}.\\
	When it is useful to distinguish, grid quantities are written in lower case Roman characters, while particle quantities are written in either upper case Roman or Greek characters. 
    Moreover, $i$ denotes grid nodes index, while $k$ denotes particles index.\\
	To discretise~\eqref{eq:weak_form}, we introduce a finite dimensional sub-space of $V$, which we call $V_h$, and a basis $\{\phi_i(x,y)\}_{i=0}^{N_h-1}$, such that $V\ni V_h = span \{\phi_i(x,y) \}_{i=0}^{N_h-1}$. Then, we project equation~\eqref{eq:weak_form} onto $V_h$, and, since $V_h$ is finite dimensional, it is possible to express any function in terms of finite sum, e.g.
	\begin{equation}\label{eq:vh}
		\mathbf{\dot v}_h (x,y) = \sum_{j=0}^{N_h -1} \mathbf{\dot v}_j \phi_j(x,y).
	\end{equation}
	Moreover, since the semi-discrete~\eqref{eq:weak_form} has to hold for any $\mathbf w_h \in V_h$, we can choose as test function any basis function. We thus get the semi-discrete weak formulation\footnote{By taking both $\mathbf{v_h}, \mathbf{w_h} \in V_h \subset [H_0^1(\Omega)]^2$, as customary in Galerkin method, we are implicitly assuming homogeneous Dirichlet boundary conditions. When this is not the case, it is enough to introduce a lifting operator, see e.g.~\cite{BrezziFortin}.}:\\
	Find $\mathbf{v_h} \in V_h$ such that $\forall i = 0:N_h -1$
	\begin{equation}\label{eq:semi-discrete}
		\int_\Omega \rho \left(\sum_{j=0}^{N_h -1} \mathbf{\dot v}_j \phi_j\right) \phi_i d\Omega = \int_\Omega (p+q)\nabla\phi_i d\Omega
	\end{equation}	
	Integrating numerically~\eqref{eq:semi-discrete} with a quadrature rule having particles volumes as weights and particles position as integration points, we can express it in algebraic form:
	\begin{equation}\label{eq:algebraic}
		\forall i = 0\ldots N_h -1 \quad	\sum_{j=0}^{N_h -1} m_{ij} \mathbf{\dot{v}}_j = \mathbf{f_i^{\,int}}
	\end{equation}
	having introduced the mass matrix $[m_{ij}]$ s.t.
	\begin{equation}\label{eq:mass_matrix}
		m_{ij}=\sum_{k=0}^{N_p-1} M_k \phi_i(X_k, Y_k)\phi_j(X_k, Y_k)
	\end{equation}
	and the internal force vector
	\begin{equation}
		\mathbf f_i=\sum_{k=0}^{N_p-1}(P_k+Q_k)\frac{M_k}{\rho_k}\nabla\phi_i(X_k,Y_k)
	\end{equation}
	where $k$ is the particles index, $N_p$ the particles number, $M_k$ is the $k-th$ particle mass such that its density is $\rho_k = \tfrac{M_k}{V_k}$, $V_k$ being its volume used as quadrature weight, $P_k$ and $Q_k$ are the $k-th$ particle pressure and artificial pressure, $\phi_i(X_k,Y_k)$ the $V_h$ base function centred on the $i-th$ grid node and evaluated at the $k-th$ particle position and $\nabla\phi_i(X_k,Y_k)$ the evaluation at $(X_k,Y_k)$ of the gradient of $\phi_i(x,y)$.\\
	In order to avoid to solve a linear system, as customary in MPM~\cite{Sandia,SteffenThesis}, we lump the mass matrix and define the nodal mass matrix whose diagonal terms are
	\begin{equation}\label{eq:lumped_mass_matrix}
		m_i = \sum_{k=0}^{Np-1}M_k\phi_i(X_k,Y_k)
	\end{equation}
	We thus have obtained an algebraic system in the form
	\begin{equation}\label{eq:algebraic2}
		\forall i = 0\ldots N_h -1 \quad	m_i \mathbf{\dot{v}}_i = \mathbf{f_i^{\,int}}
	\end{equation}
	where $[m_{i}]$ is the matrix of Eulerian grid nodal mass, which corresponds to the mass-lumped version of the original MPM algorithm mass matrix~\cite{SulskyChenSchreyer1994,Sandia,Steffen}, and $\mathbf{\dot{v}}_i$ is the unknown vector of nodal acceleration. More precisely, the expressions just obtained for $m_i$ and $\mathbf{f_i^{\,int}}$  are respectively equations (3.1) and (3.3) in~\cite{Sandia}, assuming that our model~\eqref{eq:euler} is taken into account.\\
	
	\noindent The so defined $m_i$ and $\mathbf{f_i^{\,int}}$ expressions constitute examples of what we call P2G and P2GD kernels, i.e. mapping operations from the Lagrangian particles to the Eulerian grid, where we the D stands for the derivative operator applied to the basis function, coming from having moved it from the stress to the test function in the weak formulation.\\
	As anticipated, to better analyse, measure and optimise the computational performance of the algorithm, it is important to define such mapping operations in general terms, to be able both to stress their computational features and to apply the very same kernels to different quantities as specified in the algorithm of~\cite{Sandia}.\\
	In order to do so, denote by $ \mathbf{a} = [a_i] \in \mathbb{R}^{N_h} $ the array of values of some quantity $a(x,y)$ at the grid vertices $(x_i,y_i), i=0\ldots N_h-1$, and let $ \mathbf{A} = [A_k] \in \mathbb{R}^{N_p}$ be the array of values of some quantity $A(X,Y)$ at the material points $(X_k,Y_k), k=0\ldots N_p-1$.
	\noindent It is useful to generalise the kernels to the case where the mapped quantity is not stored in memory but computed on the fly as a function of other state variables, as in equation~\eqref{eq:algebraic}. \\
	To do so, let $\mathcal{F} :\mathbb R^m \to \mathbb {R}$ and $\mathcal{G} :\mathbb R^m \to \mathbb {R}^2$; we define Particle-to-Grid and Grid-to-Particle kernels as follows:
		\begin{multline}\label{eq:p2g_gen}
			\mathrm{P2G}( \mathcal{F}, \mathbf{A}^{(0)}, \ldots, \mathbf{A}^{(m-1)};\  x_i,  y_i) := \\ 
			\sum_{k=0}^{Np-1} \mathcal{F}\left({A}_k^{(0)}, \ldots {A}_k^{(m-1)}\right)\ \phi_i(X_k,Y_k), \;  i = 0\ldots N_h-1
		\end{multline}
		\begin{multline}\label{eq:g2p_gen}
			\mathrm{G2P}(\mathcal{F}, \mathbf{a}^{(0)}, \ldots, \mathbf{a}^{(m-1)};\  X_k,  Y_k):= \\ 
			\sum_{i=0}^{N_h-1} \mathcal{F}\left({a}_i^{(0)}, \ldots {a}_i^{(m-1)}\right)\ \phi_i(X_k,Y_k),  \;  k = 0\ldots N_p-1
		\end{multline}
		\begin{multline}\label{eq:p2gd_gen}
			\mathrm{P2GD}(\mathcal{G}, \mathbf{A}^{(0)}, \ldots, \mathbf{A}^{(m-1)};\ x_i,y_i) :=\\ \sum_{k=0}^{Np-1} \mathcal{G}\left({A}_k^{(0)}, \ldots, {A}_k^{(m-1)}\right) \cdot \nabla \phi_{i}(X_k,Y_k), \; 			  i = 0\ldots N_h-1
		\end{multline}
		\begin{multline}\label{eq:g2pd_gen}
			\mathrm{G2PD}(\mathcal{F}, 
			\mathbf{a}^{(0)}, \ldots, \mathbf{a}^{(m-1)};\ X_k,Y_k):= \\
			\sum_{i=0}^{N_h-1} \mathcal{F}\left({a}_i^{(0)}, \ldots {a}_i^{(m-1)}\right)  \nabla \phi_{i}(X_k,Y_K),\;   k = 0\ldots N_p-1.
		\end{multline}
		With these definitions, the motion equation can be solved on the Eulerian grid starting from Lagrangian particles quantities, assuming that proper boundary and initial condition are applied. Indeed, equation~\eqref{eq:algebraic2} can be formulated employing the above defined general kernels~\eqref{eq:p2g_gen},~\eqref{eq:p2gd_gen} as:
		\begin{align}\label{eq:mxa=f}
			&\forall i = 0\ldots N_h-1, \nonumber\\ 
			&\mathrm{P2G}(\mathbb{I}, \mathbf M; x_i, y_i) 
			(\dot v_x)_i = 
			\mathrm{P2GD} (\zeta,{\mathbf{P,Q,M,\bm\rho}}; x_i,y_i)\\
			&\mathrm{P2G}(\mathbb{I}, \mathbf M; x_i, y_i) 
			(\dot v_y)_i = 
			\mathrm{P2GD} (\eta,{\mathbf{P,Q,M,\bm\rho}}; x_i,y_i)\nonumber
		\end{align}
		where:
		\begin{align*}
			&\mathbb I:\mathbb{R} \rightarrow \mathbb{R},\,\mathbb I(x)= x \\ 
			&\zeta:\mathbb{R}^3\times\mathbb{R}_0 \rightarrow \mathbb{R}^2,\,
			\zeta(x,y,z,t) = \begin{pmatrix}
									x+y\\
									0		
								\end{pmatrix}\frac{z}{t}\\
			&\eta:\mathbb{R}^3\times\mathbb{R}_0 \rightarrow \mathbb{R}^2,\,
			\eta(x,y,z,t) = \begin{pmatrix}
				0\\
				x+y
			\end{pmatrix}\frac{z}{t}
		\end{align*}
In the next section the detailed implementation of the algorithm is illustrated, together with the expression of computational steps according to the kernels formalism, that allow to obtain the original MPM equation as in~\cite{Sandia}\\
For the numerical discretisation of the Eulerian aspects of the method, we choose a Cartesian tensor product grid of quadrilateral elements. Among the possible choices, this one lends itself naturally to computationally efficient refinements through a quad-tree approach~\cite{quadtree,Sandia}. Moreover, we opt for $\mathbb Q_1$ FEM basis functions for the functional spaces of both the solution and the test function, see e.g.~\cite{BrezziFortin}.\footnote{In section~\ref{sec:concl} we discuss the extension to higher order basis functions.}\\
This approach corresponds to bi-linear weighting function of standard MPM,  
and is known to cause  cell-crossing error and spurious oscillations 
in the computed quantities. The reason can be easily understood by the above equations: when $\phi$ is (bi)-linear, $\nabla \phi$ is constant, but potentially different in neighbouring cells. Thus, when crossing cells, particles might experience discontinuous oscillations. This issue can be numerically tackled with many strategies, among which adding artificial viscosity~\cite{MPMBook} or choosing higher order basis functions. For simplicity of implementation, we choose the first, leaving the second for a future work.\\
With respect to standard MPM~\cite{YorkThesis,Sandia}, the present algorithm adds a step to take into account the flow around obstacles. Namely, we subtract the inward component of both linear momentum and force from the grid nodes whose cell is marked as an obstacle cell in following equations (stemming from the method for granular materials~\cite{normali}). \\
$\forall i = 0\ldots N_h -1$:
\begin{equation}\label{eq:normali1}
	{(\mathbf {mv})_i^o}'
	= (\mathbf {mv})_i^o -  \dfrac{(\mathbf {mv})_i^o\cdot\mathbf n_i^o - |(\mathbf {mv})_i^o\cdot\mathbf n_i^o|}{2} \mathbf n_i^o
\end{equation}
\begin{equation}\label{eq:normali2}
	{\mathbf f_i^o}' = \mathbf f_i^o -  \dfrac{\mathbf f_i^o\cdot\mathbf n_i^o - |\mathbf f_i^o\cdot\mathbf n_i^o|}{2} \mathbf n_i^o
\end{equation}
where $\mathbf{mv}$ is the linear momentum, $\mathbf f$ the internal force, the subscript denotes the $i-th$ grid node, the superscript $o$ stands for obstacle and $\mathbf n$ is the normal unit vector, positive if directed outward from the obstacle. \\
Normal unit vectors to a domain profile $\partial\Omega$ are computed as gradients of the Signed Distance Function, defined as
\begin{equation}\label{eq:sdf}
	sd(\mathbf x) = \begin{cases}
		-\inf_{\mathbf y\in\partial\Omega}d(\mathbf x, \mathbf y) \quad \mathbf x\in\Omega\\
		\inf_{\mathbf y\in\partial\Omega}d(\mathbf x, \mathbf y) \quad \mathbf x\notin\Omega
	\end{cases}
\end{equation}
where $d(\mathbf x,\mathbf y)$ is the Euclidean distance. Taking the gradient yields\footnote{An algorithm for signed distance function computation can be found in~\cite{sdf}, and an example in a machine learning based framework that implicitly represents geometries from noisy surface points can be found in~\cite{sdfmox}.}
\begin{equation}
	\nabla sd(\mathbf x)  =  \mathbf n(\mathbf x) \quad \mathbf x\in\partial\Omega
\end{equation}
\noindent Finally, we use explicit time integration~\cite{Sandia}, whose stability is subject to a CFL criterion
\begin{equation}\label{CFL}
	\Delta t < \alpha \dfrac{\Delta x_{min}}{v_{max}}
\end{equation}
where $\alpha$ collects constants.
The range of admissible time steps is bounded from above by the ratio between the smallest grid cell dimension $\Delta x_{min}$ and the highest speed $v_{max}$, considering both particles velocity and speed of sound, given by the square root of the adiabatic index $\gamma$ multiplied by the ratio of gas pressure and mass density. The efficiency cost of enforcing small time steps is balanced by the need of resolving fast dynamics. In addition, explicit-in-time methods sidestep the need to solve global linear systems at each time step, thus enjoying better scalability properties than implicit methods.

\section{Portable MPM Implementation}\label{sec:alg_and_impl}
This section describes  how we formulated the MPM algorithm of~\cite{Sandia} within the framework that was presented in section~\ref{sec:num} and how we implemented it to ensure portability on different architectures and efficiency on GPUs. The development of the code benefited in part from publicly available Open Source libraries: Thrust\footnote{\href{https://github.com/NVIDIA/cccl/tree/main/thrust}{https://github.com/NVIDIA/cccl/tree/main/thrust}}, rocThrust\footnote{\href{https://github.com/ROCm/rocThrust}{https://github.com/ROCm/rocThrust}}, ``JSON for Modern C++"\footnote{\href{https://github.com/nlohmann/json}{https://github.com/nlohmann/json}}, and quadgrid.\footnote{\href{https://github.com/carlodefalco/quadgrid}{https://github.com/carlodefalco/quadgrid}}\\
Let $\mathbb I, \zeta, \eta$ denote the fields defined in~\ref{sec:num} and let $\theta$ be:
\begin{equation*}
	\theta:\mathbb{R}\times\mathbb{R}_0 \rightarrow \mathbb{R},\,\theta(x,y)= \frac{x}{y} 
\end{equation*} 
In detail, the algorithm implementation is composed of the following steps:
\begin{enumerate}
	\item Data generation: the creation of simulations input is performed separately from code execution. For simple geometries it is handled via ad-hoc C++ scripts; for complex geometries, requiring a careful computation of the signed distance function, an Octave\footnote{\href{https://octave.org/}{https://octave.org/}} script is used instead, to take advantage of Octave built-in functions. In both cases, since  objects are considered at rest, normal unit vectors are computed only once, and boundary cells are marked with a flag corresponding to the boundary condition. Finally, the initial data file is saved in JSON format. 
	\item MPM class object construction: since the most computationally costly operations are the memory related ones, such as I/O and copies between host and device, a class object for the method is constructed only once, reading grid and particles data from the input file, data is saved on host memory and copied to device memory. So doing, is possible to construct functor objects which will be called on the device during the time loop, without the need of further copies.  
	\item The time advancing loop starts; inner steps, except from output operation, are implemented as Thrust or STL {\ttfamily for\_each}es, depending on the choice made at compile time.
	\begin{enumerate}[(i)]
		\item Grid variables are reset and current grid variables are overwritten with the ones saved during construction stage, performing a device to device copy. This step can be performed since all history dependence is maintained by Lagrangian particles.
		Specifically, the variables that have been reset to their original input values are vectors holding grid nodal mass, $x$ and $y$ components of linear momentum, force normal to the obstacle's unit vector, as well as markers of whether such node has to be considered on a domain boundary or inside an obstacle.
		\item Equations~\eqref{eq:p2g_gen} and~\eqref{eq:p2gd_gen} are applied to map particles mass, momentum and force to the grid:
		\begin{align*}
			&\forall i = 0, \ldots, N_h -1\\
			&m_i = \mathrm {P2G} (\mathbb I, \mathbf M; x_i, y_i)\\
			&(mv_x)_i = \mathrm {P2G} (\mathbb I, \mathbf{MV_x}; x_i, y_i) \\
			&(mv_y)_i = \mathrm {P2G} (\mathbb I, \mathbf{MV_y}; x_i, y_i)\\
			&(f_x)_i = \mathrm{P2GD} (\zeta,{\mathbf{P,Q,M,\bm\rho}}; x_i,y_i)\\
			&(f_y)_i = \mathrm{P2GD} (\eta, {\mathbf{P,Q,M,\bm\rho}}; x_i,y_i)
			\end{align*}

        \item Boundary conditions are enforced on grid nodes momentum and force. Specifically, at each time step and for each boundary node marked as of obstacle type, inward components of momentum and force are subtracted as in equations~\eqref{eq:normali1} and~\eqref{eq:normali2}.
		\item Grid nodes momentum is advanced in time.\footnote{In the implementation we use the linear momentum, as in eq. (3.5) of~\cite{Sandia}, instead of the equivalent formulation in terms of mass and velocity as in equation~\eqref{eq:mxa=f} or as in eq. (3.23) of~\cite{SteffenThesis}.}
		
		\begin{align*}
			&\forall i = 0, \ldots N_h -1\\
			&(mv_x)_i' = (mv_x)_i + \Delta t (f_x)_i\\
			&(mv_y)_i' = (mv_y)_i + \Delta t (f_y)_i
		\end{align*}
		
        \item Velocity and acceleration  {generalised} G2P~\eqref{eq:g2p_gen} is applied, and next, velocity is updated. Acceleration itself is not stored, but computed as the result coming from a local G2P of the relevant nodes ratio between force and mass. 
		Following standard MPM~\cite{Sandia}, we store the G2P velocity in a temporary velocity vector $\mathbf{\overline V}$ and don't use it to update particles velocity; however, we postpone its usage.
		
        \begin{align*}
			&\forall k = 0, \ldots N_p -1\\
			&(\overline V_x)_k = \mathrm {G2P} (\theta, \mathbf{mv_x}, \mathbf{m}; X_k, Y_k)\\
				&(\overline V_y)_k = \mathrm {G2P} (\theta, \mathbf{mv_y}, \mathbf{m}; X_k, Y_k)\\
				&(A_x)_k = \mathrm {G2P} (\theta, \mathbf{f_x}, \mathbf{m}; X_k, Y_k)\\
				&(A_y)_k = \mathrm {G2P} (\theta, \mathbf{f_y}, \mathbf{m}; X_k, Y_k)
			\end{align*}
		
		\item Momentum P2G~\eqref{eq:p2g_gen} is performed.
		\begin{align*}
				&\forall i = 0, \ldots N_h -1\\
				&(mv_x)_i =  \mathrm {P2G} (\mathbb I, \mathbf{MV_x}; x_i, y_i)\\
				&(mv_y)_i =  \mathrm {P2G} (\mathbb I, \mathbf{MV_y}; x_i, y_i)
		\end{align*}
		\item Boundary conditions on the updated grid values are enforced.
		\item Particles strain rate tensor (velocity gradient) is computed as the G2PD~\eqref{eq:g2pd_gen} of the ratio between grid nodes momentum and mass, and then particles properties are updated.
			\begin{align*}
				&\forall k = 0, \ldots N_p -1\\
				&\nabla (V_x)_k = \mathrm {G2PD} (\theta, \mathbf{mv_x}, \mathbf{m}; X_k, Y_k)\\
				&\nabla (V_y)_k = \mathrm {G2PD} (\theta, \mathbf{mv_y}, \mathbf{m}; X_k, Y_k)\\
				&E_k' = E_k + \Delta t \dfrac{-(P_k+Q_k)}{\rho_k} \nabla\cdot \mathbf V_k\\
				&\rho_k' = \frac{\rho_k}{1+\Delta t \nabla \cdot \mathbf V_k}\\
				&P_k' = (\gamma -1)\rho_k' E_k'
			\end{align*}
		\item Particles position are updated using $\mathbf{\overline V}$:
			\begin{align*}
				&\forall k = 0, \ldots N_p -1\\
				&(X_k)' = X_k + \Delta t (\overline{V_x})_k\\
				&(Y_k)' = Y_k + \Delta t (\overline{V_y})_k
			\end{align*}
		Thus, the time advancing scheme is the same of~\cite{Sandia}. However, particles moving is postponed w.r.t. the standard algorithm~\cite{Sandia}, so that, in all the above steps, basis functions and their gradient can be computed on the fly, without the need to store the value they had at former particles position.
		\item Particles-to-grid connectivity map is updated with new positions - this step is computationally efficient also thanks to the choice of a Cartesian grid.
		\item Only if the initial data options ask for $N$ intermediate outputs, a device to host memory copy and an output from host to disk are performed at $N$ moments equidistant in time.
		\item  Optionally, a re-ordering algorithm is applied to keep data locality, i.e. to keep contiguous particles in the physical simulation also close in memory. The algorithm is based on an indexing vector built iterating over the structure that stores information about which particles are contained in every grid cell, such that the n-th element of the ordering vector indicates the current sorting particle index which has to be at n-th position after re-ordering.
		\item A control on the CFL condition~\eqref{CFL} is enforced, taking into account particles velocity (momentum over mass and $\mathbf{\overline V}$) and the speed of sound.
		\item Time step is updated, loop ends.
	\end{enumerate}
	\item A final device-to-host memory copy and output is made.
\end{enumerate}

It is worth noting that, even though the sum in the analytical expression of the Grid-to-Particle kernels is performed on the grid index, in the implementation also G2P and G2PD kernels are performed on particles index, and, internally, they rely on the particle-to-grid map, as shown in the snippet in Fig.~\ref{fig:g2ptemplate}. Indeed, the implementation has been found to be more efficient when the Lagrangian character of the method is taken into account. In particular, in the parallelisation scheme one thread is requested for each particle.\\
Algorithm~\ref{algo1} summarises the for loop procedure, with a simplified notation.
\begin{algorithm}[ht]
	\caption{Time loop sketch}\label{algo1}
	\begin{algorithmic}[1]
		\While{ $t < t_f$}
		\State Grid reset
		\State Mass, momentum and force P2G 
		\[\qquad m = \mathrm{P2G} (M),\ \mathbf {mv} = \mathrm{P2G} (M\mathbf V),\ \mathbf f = \mathrm{P2GD} ((P+Q) \tfrac{M}{\rho})\]
		\State Boundary conditions enforcement on $\mathbf {mv},\,\mathbf f$ \& obstacle treatment as in~\eqref{eq:normali1},~\eqref{eq:normali2}
		\State Momentum equation solution on the grid
		\[ \qquad(\mathbf {mv})'=\mathbf {mv} + \Delta t \mathbf f\] 
		\State G2P mapping 
		\[\qquad \overline{\mathbf V} = \mathrm{G2P}(\tfrac{\mathbf {mv}}{m}),\ {\mathbf A} = \mathrm{G2P}(\tfrac{\mathbf f}{m})
		\]
		\State Velocity update
		\[\qquad\mathbf V' = \mathbf V + \Delta t \cdot \mathbf A\]
		\State Momentum P2G
		\[\qquad \mathbf {mv} = \mathrm{P2G} (\mathbf {MV})\]
		\State Boundary condition enforcement on $\mathbf {mv}$
		\State Particles properties update:
		\begin{align*}
			&\qquad\nabla \mathbf V = \mathrm{G2PD}(\tfrac{m\mathbf v}{m})\\
			&\qquad E' = E + \Delta t \dfrac{-(P+Q)}{\rho} \nabla\cdot \mathbf V\\
			&\qquad\rho' = \dfrac{\rho}{1+\Delta t \nabla \cdot \mathbf V}\\
			&\qquad P' = (\gamma -1)\rho' E'
		\end{align*}
		\State Particles moving $\mathbf X' = \mathbf X + \Delta t \overline{\mathbf V}$
		\State  Optional particles re-ordering
		\State  CFL condition~\eqref{CFL} check \& time advance
		\EndWhile
	\end{algorithmic}
\end{algorithm}

\noindent This algorithm is of special interest for acceleration on GPUs since it can be written in terms of the four kernels~\eqref{eq:p2g_gen}-\eqref{eq:g2pd_gen}, which present interesting parallelisation potentialities. Grid-to-Particles operations are intrinsically Single Instruction Multiple Data (SIMD) parallelisable, since data is mapped from grid nodes to particles, and thus particles are independent from one another during writing stage. Particles-to-Grid kernels present the same potentialities, however, if they are implemented strictly following their analytical definition, they present a data race: different particles need access in both read and write mode to the same grid nodes.\\
In this work, we chose to patch the Particles-to-Grid and Grid-to-Particles kernels written in the external quadgrid library to adapt them to GPU execution via Thrust rather than writing ad hoc versions in the code, at the expense of some further optimisation, such as specific kernel-fusions, that could have been added by exploiting the peculiarities of our model~\eqref{eq:fullmodel}, to arrive at more general and portable conclusions.  This way, there is only one computational kernel per operation~\eqref{eq:p2g_gen}-\eqref{eq:g2pd_gen}. In any case, preliminary tests have shown that applying a kernel-fusion approach~\cite{Thrust} where possible wouldn't have affected the execution time order of magnitude. \\

\noindent{To better illustrate the concepts introduced in the last 
	paragraph, we sketch and discuss below the implementation and use of the 
	two main kernels in the algorithm, namely the Grid-to-Particle (G2P) 
	operation~\eqref{eq:g2p_gen} and the Particle-to-Grid operation~\eqref{eq:p2g_gen}.
	A simplified sketch of the implementation of the former 
	is given in Fig.~\ref{fig:g2ptemplate}, while the latter is described in
	Fig.~\ref{fig:p2gtemplate}; the reported code corresponds roughly to the 
	actual implementation available in the quadgrid library except for a few 
	details that are left out only to avoid distraction, as they are not 
	interesting for the current discussion. As already noted, the G2P operation 
	is easily implemented via a loop over all particles and lends itself quite
	naturally to SIMD parallelisation as all write operations are independent
	and may be carried out concurrently. The actual looping is performed via
	the 
 {\ttfamily for\_each}
 method, which can be used to perform any operation
	on any number of variables associated to a particle by using a \emph{counting
		iterator} and a suitably constructed \emph{callable object} as shown in 
	Fig.~\ref{fig:g2pforeach} and Fig.~\ref{fig:g2pcallable}. The P2G operation
	is implemented along lines very similar to those followed for G2P. Specifically,
	as shown in Fig.~\ref{fig:p2gtemplate} the looping is done over the list of particles
	just as for G2P, which is made possible by the index array {\ttfamily ptcl\_to\_grid}
	that lists, for each particle, the grid cell within which it is located. Efficiently 
	building the latter data structure heavily depends on the use of structured grid.
	The main difference between P2G and G2P consists in the need, in the former,
	to work around a data race
	that is created when threads working on different particles try to write data related
	to the same neighbouring cell vertex. The simplest approach, followed in this work, is
	to make the write operations \emph{atomic} by 
	using {\ttfamily atomicAdd}
 (as shown for sake of example in Fig.~\ref{fig:p2gtemplate}),
	which is readily available in CUDA and HIP but can also be re-implemented in standard C++
	based on the 
 {\ttfamily std::atomic\_ref}
 class template introduced in the C++20 standard.
	Although our tests show that the use of atomics does not seem to lead to much of
	a performance degradation on GPUs, we discuss in a later section a possible alternative based
	on data reordering and mesh colouring that can be used to avoid atomics.
}

\begin{figure}
	\begin{lstlisting}[style=cpp, firstnumber=1, numbers=left, numberstyle=\tiny,showtabs=false]
		template<class callable_t>
		G2P (idx_t iptcl, callable_t *grid_variable, 
		ptcl_variable_t *ptcl_variable) {
			
			idx_t icell = ptcl_to_grid[iptcl];
			real_t x = ptcl_x[iptcl];
			real_t y = ptcl_y[iptcl];
			
			for (idx_t inode = 0; inode < 4; ++inode) {
				real_t N = shape_function (x, y, inode, icell);
				idx_t ginode = local_to_global (icell, inode);
				
				ptcl_variable[iptcl] += 
				N * (*grid_variable)(ginode);
			}
			
		};
	\end{lstlisting}
	\caption{{Example template function to demonstrate 
			the approach used to implement the G2P kernel. 
			This is not the actual code used but a
			conceptualised, reduced example to explain the basic idea.
			In particular note that in this simplified snippet (and some of the 
			following) we use, for sake of simplicity of presentation, standard
			pointers, while the
			actual implementation uses some form of iterator.}}
	\label{fig:g2ptemplate}
\end{figure}

\begin{figure}
	\begin{lstlisting}[style=cpp, firstnumber=1, numbers=left, numberstyle=\tiny,showtabs=false]
		struct x_velocity_t {
			grid_data_t *gdata = &grid_data;
			real_t operator()(idx_t ginode) {
				real_t vx = gdata -> get (ginode, "x_momentum");
				real_t m  = gdata -> get (ginode, "mass");
				return vx / m;
			}
		} *x_velocity;
		
		x_velocity_g2p = 
		[x_velocity, ptcl_x_velocity] (idx_t iptcl) {
			G2P (iptcl, x_velocity, ptcl_x_velocity);
		};
	\end{lstlisting}
	\caption{{Instantiation of a callable object that 
			implements grid to particle interpolation of the 
			$x$--component of the velocity. 
			This is not the actual code used but a conceptualised, reduced
			example to explain the basic idea. In the code above, the purpose of 
			the callable object pointed to by {\ttfamily x\_velocity} is to 
			compute \emph{on--the--fly} a function of the state variables at 
			a grid vertex, so that its value can be interpolated at the 
			particle locations without requiring extra storage or data 
			transfers.}}
	\label{fig:g2pcallable}
\end{figure}

\begin{figure}
	\begin{lstlisting}[style=cpp, firstnumber=1, numbers=left, 
		numberstyle=\tiny,showtabs=false]
		counting_iterator<idx_t> first_p(0); 
		ounting_iterator<idx_t> last_p(num_particles);
		for_each (execution_policy, 
		first_p, last_p, x_velocity);
	\end{lstlisting}
	\caption{{Example of performing an actual G2P 
			operation by iterating over all particles by means of the
			{\ttfamily for\_each} method and counting iterators.}}
	\label{fig:g2pforeach}
\end{figure}

\begin{figure}
	\begin{lstlisting}[style=cpp, firstnumber=1, numbers=left, numberstyle=\tiny,showtabs=false]
		template<class callable_t>
		P2G (idx_t iptcl, ptcl_variable_t *grid_variable,
		callable_t *ptcl_variable,) {
			
			idx_t icell = ptcl_to_grid[iptcl];
			real_t x = ptcl_x[iptcl];
			real_t y = ptcl_y[iptcl];
			
			for (idx_t inode = 0; inode < 4; ++inode) {
				real_t N = shape_function (x, y, inode, icell);
				idx_t ginode = local_to_global (icell, inode);
				
				atomicAdd (grid_variable + iptcl, 
				N * (*ptcl_variable)(ginode));
			}
			
		};
	\end{lstlisting}
	\caption{{Example template function to demonstrate 
			the approach used to implement the G2P kernel. 
			This is not the actual code used but a
			conceptualised, reduced example to explain the basic idea.
			In particular note that in this simplified snippet (and some of the 
			following) we use, for sake of simplicity of presentation, standard
			pointers, while the actual implementation uses some form of iterator.}}
	\label{fig:p2gtemplate}
\end{figure}


\section{Benchmarks}\label{sec:res}

\noindent In this section we show the results of 2D numerical simulations of gas dynamics obtained with the described implementation of MPM, with a focus on transonic and supersonic gas flow over solid obstacles. A first general evaluation of accuracy in subsection~\ref{subsec:benchmarks} is followed by an assessment of the sensitivity of the results to the choice of numerical parameters in subsection~\ref{subsec:sensitivity};
{
	in particular, the effects of numerical viscosity components on mass density are shown for a test case with analytical solution in subsection~\ref{subsec:sensitivity}.\\
	In subsection~\ref{subsec:perf} we then test the efficiency and performance of the implementation in providing parallel scalability on NVIDIA A100 GPUs. Lastly, in subsection~\ref{subsec:portab}, we assess the performance portability of the developed code to different CPUs and GPUs architecture, which is our main claim.\\
	The HPC experiments were carried out on NVIDIA A100 SXM4 40 GB GPUs, Intel Xeon Platinum 8260 and AMD EPYC Rome 7402 CPUs at Leonardo S.p.A.'s in-house HPC facility \textit{davinci-1}, and on AMD EPYC 7313, AMD EPYC 7413 CPUs and AMD Radeon Instinct MI210 GPUs at MOX Laboratory for Modeling and Scientific Computing at Politecnico di Milano's Mathematics Department.\\
	All tests are performed in double precision, i.e. using FP64.}
 {Depending on the architecture, better performance could be achieved with lower precision floating points; for example, the NVIDIA A100 GPUs employed here should gain a $2\times$ factor in computational time when working with FP32. However, although the code supports different choices, we have focused on doubles only, since floats could lead to numerical difficulties in some test cases, and since other GPUs, such as the AMD MI 210, are able to handle double and single precision floats without any theoretical performance degradation.}

\subsection{Test cases}\label{subsec:benchmarks} 
{\noindent
	In the following numerical tests, }we adopt non-dimensional units as in~\cite{Sod}, and we model the fluid as a perfect, bi-atomic gas, with specific heat ratio:
\begin{equation}
	\gamma = \dfrac{c_p}{c_v} = 1.4
\end{equation}
Moreover, except in the Sod Shock Tube, in the far field away from any obstacle, we set fluid density and pressure as:
\begin{itemize}
	\item $\rho_\infty = 1.4$;
	\item $p_\infty = 1$;
\end{itemize}
where the subscript $\infty$ denotes unperturbed conditions. Thus, we obtain a unitary speed of sound:
\begin{equation}
	c_{s,\infty} = \sqrt{\gamma\ \dfrac{p_\infty}{\rho_\infty}} = 1
\end{equation} 
and an unperturbed Mach number equal to the reference speed:
\begin{equation}
	M_\infty = \tfrac{v_\infty}{c_{s,\infty}} = v_\infty
\end{equation}
For time integration, in our code we enforce the time step to be:
\begin{equation}
	\Delta t \leq \dfrac{1}{2} \dfrac{h_{min}}{v_{max}+c_{s,max}}
\end{equation}
This requirement is more conservative than necessary, and, in the limits of the CFL condition, could be partially relaxed. However, we choose it to guarantee that particles can't move across more than one grid cell per time iteration; so doing, in the framework of our algorithm, we avoid the risk that particles pass through any boundary, and ensure that their motion will respect boundary conditions.\\
In all test cases except in the Sod Shock Tube, initial particles position is sampled by a uniform random distribution in the fluid domain. Indeed, a sufficient randomness of particles position, especially in particles recycling from outlet to inlet, is fundamental to obtain stable and physically meaningful simulations, in agreement with results in the literature on Particle-In-Cell methods in plasma physics~\cite{sonnendrucker}. On longer runs (i.e. above tens of physical seconds) a symplectic time integration scheme might be useful~\cite{sonnendrucker,Berzins}.\\
In the following, we denote with $ppc$ the average number of fluid particles per fluid cell, with $N_C$ the total number of cells, and with $N_P$ the total number of particles.

{
	\paragraph{Sod Shock Tube}
	In the Sod Shock Tube, two compressible fluids, initially at rest, are divided by a diaphragm. The first fluid has $\rho_1 = 1$ and $p_1=1$, the second $\rho_2=0.125$ and $p_2=0.1$. At time $t=0$ the diaphragm is removed, and the system evolution is studied; the following phenomena are expected:
	\begin{itemize}
		\item the two initial zones shrink progressively towards domain borders;
		\item a rarefaction wave propagates toward the high density zone;
		\item in the position where the diaphragm was, which now is moving towards low densities, a contact discontinuity develops; 
		\item a shock wave propagates towards low density zones. 
	\end{itemize}
        At time $t=0.143$ we stop the simulation, thus particles do not leave the domain. In order to reproduce the 1D problem, we introduce a mesh of $1\times100$ cells; moreover, to be able to compare the MPM solution with the reference one, we also set initial particle position as in~\cite{York,YorkThesis}.\\
	In Figure~\ref{fig:sod_cfr_toro} we show the final particles position, together with a comparison of the computed solution, its linear interpolation on the grid, and the exact solution computed following the procedure shown in~\cite{Toro}. Results are in very good agreement with the MPM literature~\cite{York,AdVreview}.
}	
\begin{figure*}[h]
		\centering
		\includegraphics[width=\linewidth]{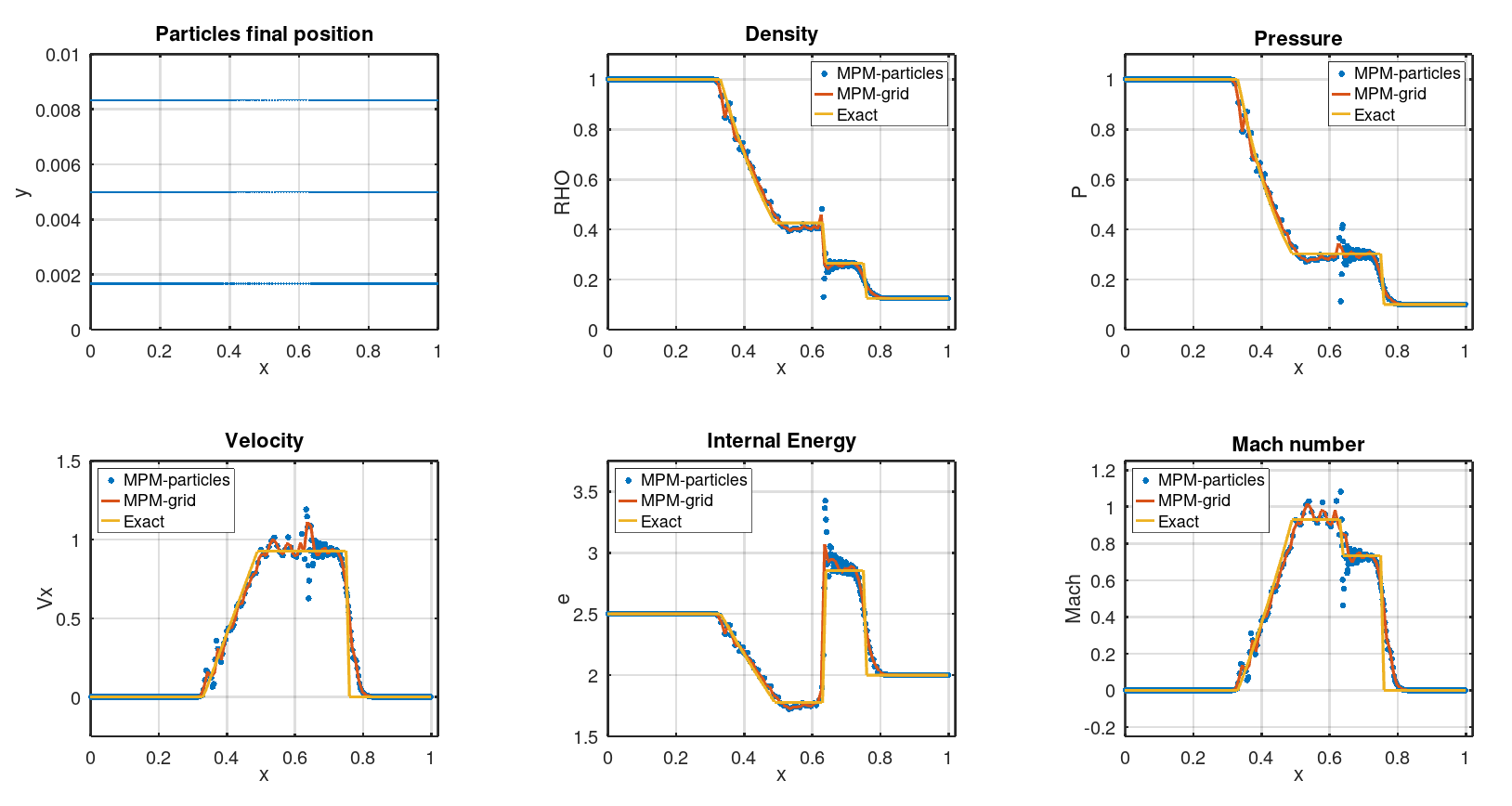}
		\caption{Sod Shock Tube: comparison of MPM solution computed on particles, interpolated on the grid, and exact solution computed as in \cite{Toro}}
		\label{fig:sod_cfr_toro}
	\end{figure*}

\paragraph{Supersonic flow past a cylinder}
We then consider the case of supersonic, Mach 3 flow past a cylinder of radius $0.25$ and centre $(0.6,1)$ in a $[0,4]\times[0,2]$ rectangular domain.
We enforce wind tunnel--like boundary conditions, with inflow conditions at the left vertical boundary, outflow conditions at the right vertical boundary, and slip conditions at the top and bottom boundaries, as in~\cite{MaierLike}.\\
Specifically, we set:
\begin{equation}\label{eq:bc}
	\begin{cases}
		\mathbf v_{inflow} = (3,0)\\
		\mathbf v_{top} \cdot \mathbf n = \mathbf v_{bottom} \cdot \mathbf n = 0
	\end{cases}
\end{equation}
where $\mathbf n$ denotes the unit normal vector. A particles-recycling approach is adopted, whereby particles leaving the domain at the outflow boundary are regenerated in a random cell at the inflow boundary with inflow conditions, keeping the total number of particles constant throughout the simulation.
In this test case, as gas particles travel past the cylinder, a shock wave develops at the front and two oblique shocks in the rear. The first shock is reflected by walls and interacts with the back shocks; see Figure~\ref{fig:cylinder9} and its caption for numerical details.\\

\begin{figure*}
	\centering
	\includegraphics[width=\linewidth]{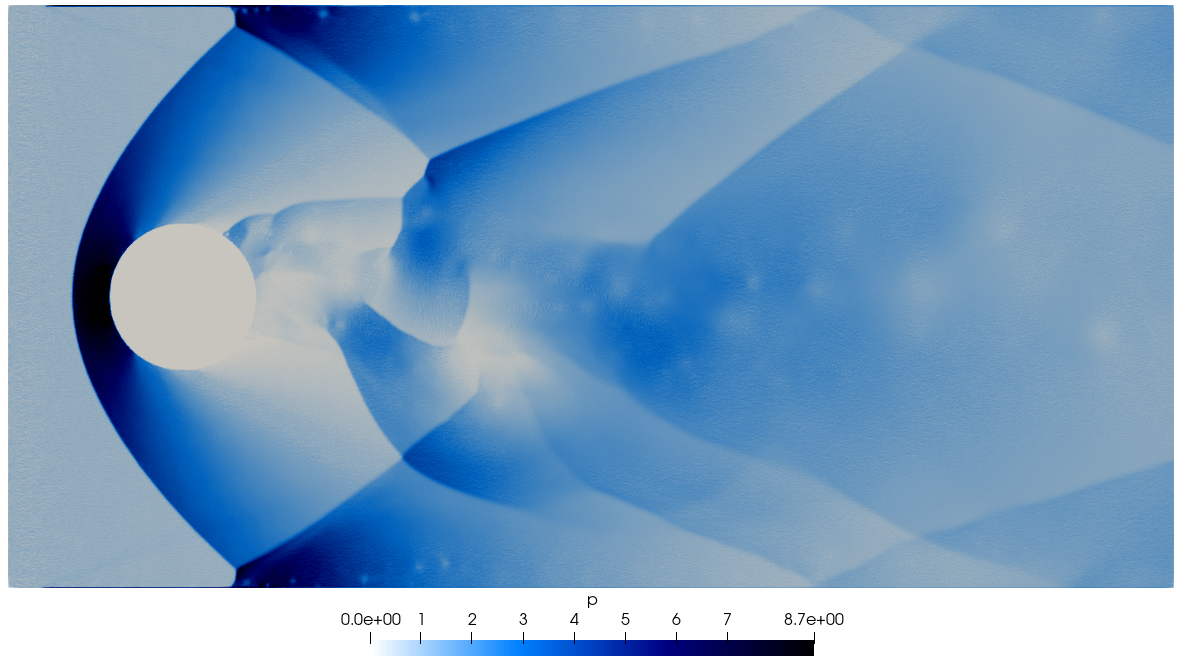}
	\caption{Mach 3 flow past a cylinder: pressure at $t=10\,s$, $N_C= 10^6$, $ppc = 16$, $N_P = 15.6 \cdot 10^6$}
	\label{fig:cylinder9}
\end{figure*}

The MPM implementation is able to correctly simulate the main flow features; computed shock-waves angle and flow deflection angle (Figure~\ref{fig:cylinder9}) are found to be in agreement with experimental references~\cite{vandyke} as well as with results which can be obtained with other codes, e.g., deal.II, when the same polynomial order is used for basis functions. {In particular, the cylinder test case is simulated in the deal.II documentation~\cite{matthias_maier_2020_3698223}\footnote{\href{https://www.dealii.org/current/doxygen/deal.II/step\_69.html\#Results}{\scriptsize https://www.dealii.org/current/doxygen/deal.II/step\_69.html\#Results}}. With our implementation, we are able to reproduce the main features of the front shock starting from a very low resolution, with details, especially in the rear, getting better as resolution increases; more details are shown in subsection~\ref{subsec:sensitivity}. Overall, the results are in agreement with~\cite{matthias_maier_2020_3698223}, while the lower level of detail in comparison with some results in the literature is likely due to the first order space discretisation used here (\cite{CLAYTON2023111926,ryujin,MaierLike}). 
	
	\paragraph{Transonic flow past an aerofoil} Next, we test the method in simulating compressible, transonic flow past an OAT15a aerofoil at Mach 0.73, to assess the ability of the implementation in dealing with more complex-shaped obstacles.
	The signed-distance function approach, in conjunction with the implemented obstacles treatment~\eqref{eq:normali1},\eqref{eq:normali2}, naturally fits the aerofoil profile (Figure~\ref{fig:sdf2}). Since, rather than using an analytic expression, the profile and its normal unit vectors are generated reading from a file containing a set of points, the method proves to be sufficiently general and not bounded to particular geometric shapes.\\	
	Within the $[0,0.4]\times[0,0.27]$ domain, we enforce inflow conditions at the left boundary, with 
	\begin{equation}
		\mathbf{v}_\infty = (0.73, 0)
	\end{equation}
	and outflow conditions at other boundaries.
	To test the physical reliability of the approach, we choose a large attach number to emphasise dynamical effects and display velocity magnitude and streamlines (Figure~\ref{fig:oat15a}). \\
	These are found to be in line with what is expected from physical considerations: we can see airflow speeding up as it approaches the aerofoil, with a higher velocity magnitude in the top area with respect to the bottom one.

 \begin{figure}
	\centering
	\includegraphics[width=\linewidth]{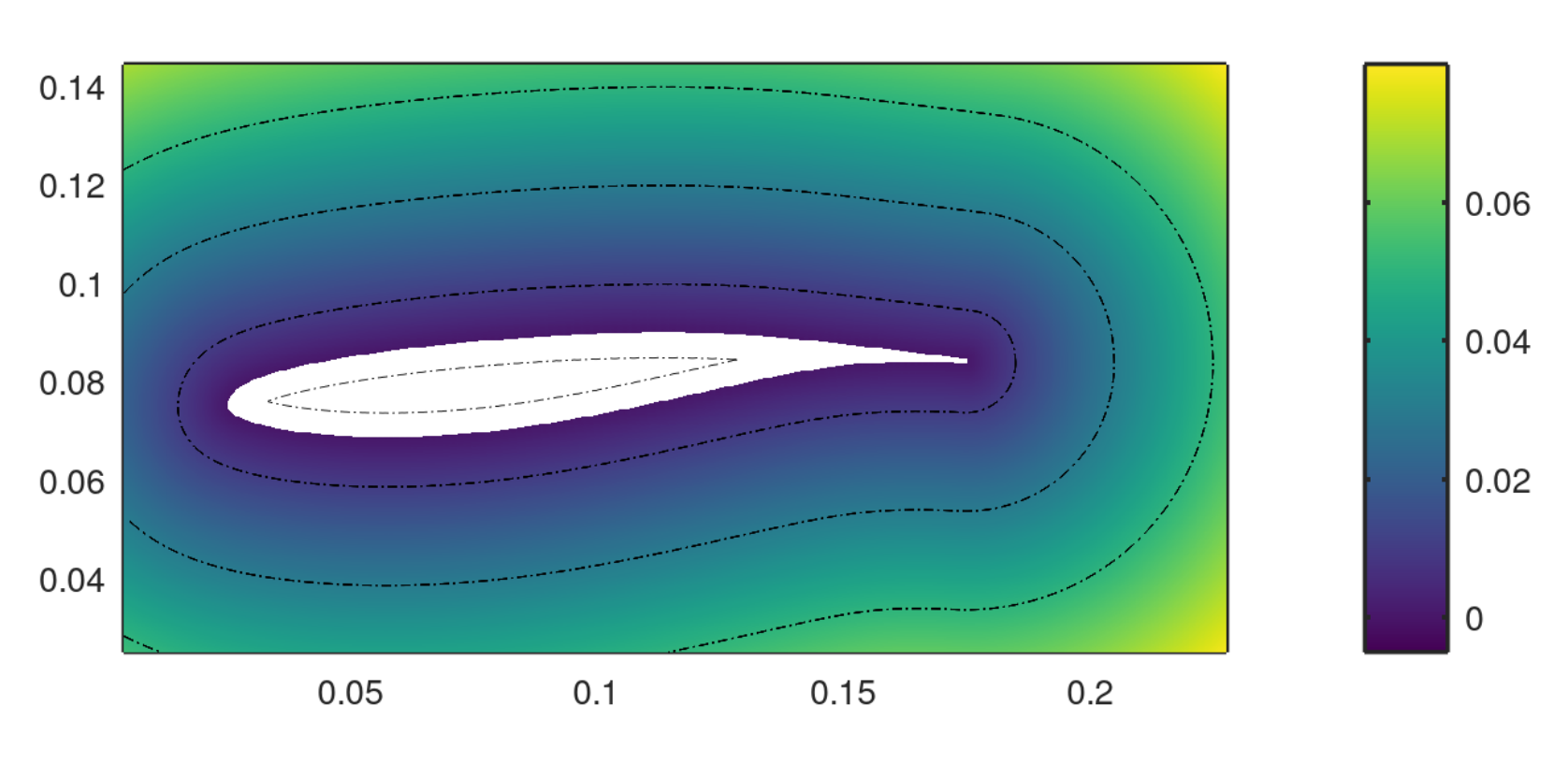}
	\caption{Signed distance function for OAT15a aerofoil at $\alpha = -3.5^\circ$}\label{fig:sdf2}
\end{figure}

\begin{figure}
	\centering
	\includegraphics[width=\linewidth]{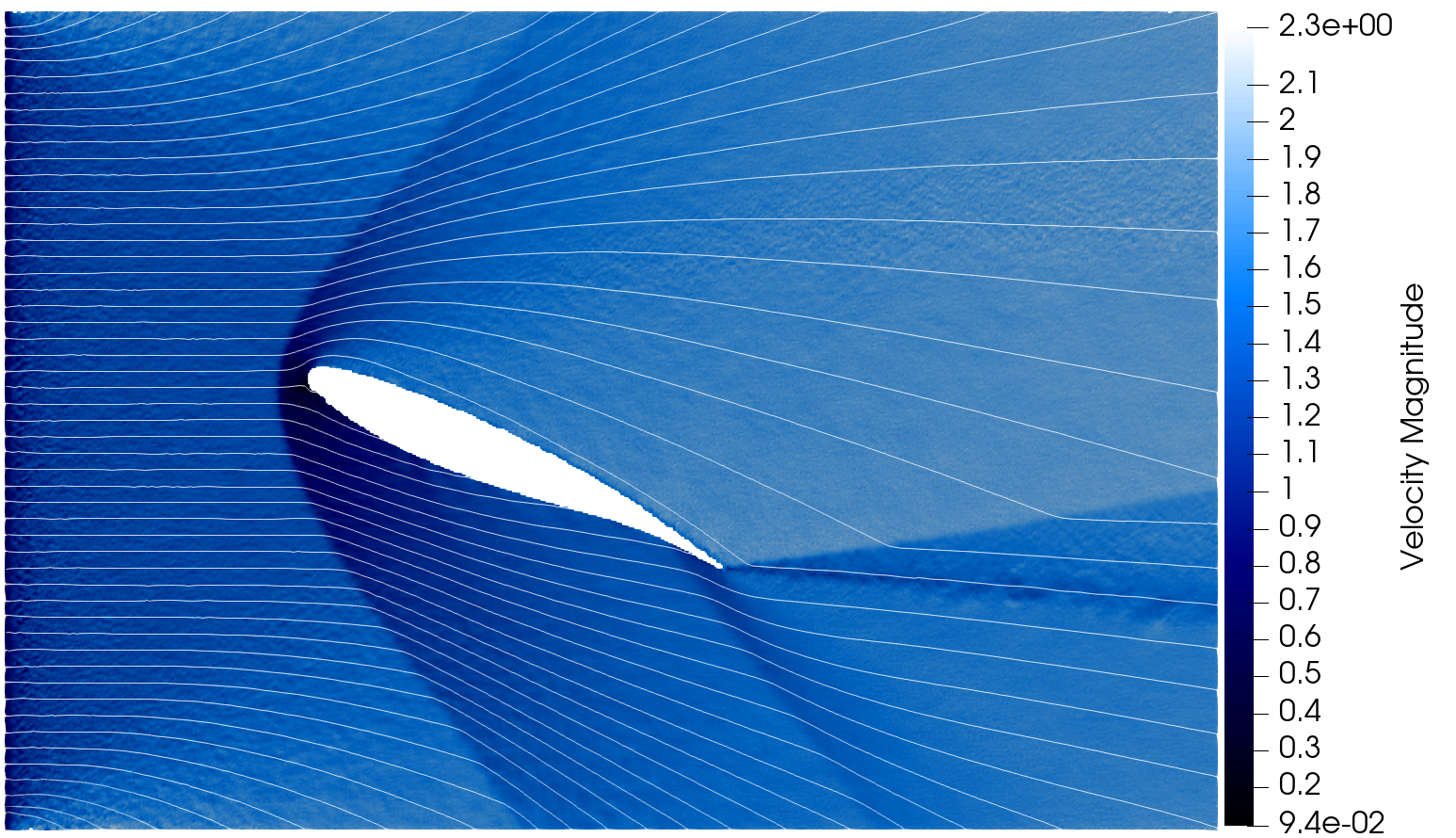}
	\caption{OAT15a Mach 0.73 flow: velocity streamlines at $\alpha = 25^\circ$, $t=1s$, $N_C = 247\cdot10^3$, $ppc = 9$, $N_P = 2.18\cdot10^6$.}\label{fig:oat15a}
\end{figure}
	
	
	\subsection{Sensitivity to numerical parameters}\label{subsec:sensitivity}
	{
		\paragraph{Artificial viscosity}
		Artificial viscosity introduced in section~\ref{sec:num} is composed of two main terms: a quadratic one and a linear one. The first, introduced by von Neumann-Richtmyer~\cite{mu_art_1}, is a viscous term that, being added to the pressure in the vicinity of shock waves, reduces numerical oscillations and enables the capturing of the shock. The second, introduced by Landshoff~\cite{mu_art2}, is a linear term added to the quadratic one to act specifically in the region behind the shock front. Further details on the application of this viscosity to MPM can be found in~\cite{MPMBook}.\\
		In this paragraph we study the effect of the artificial viscosity, comparing the final density obtained enforcing $c_0=1, c_1 = 1$ in equation~\eqref{eq:muart} and shown in Figure~\ref{fig:sod_cfr_toro}, with the one obtained by:
		\begin{itemize}
			\item turning off the artificial viscosity;
			\item applying only the von Neumann-Richtmyer term, enhanced by a factor 10;
			\item applying only the Landshoff term, enhanced by a factor 10.
		\end{itemize}
		Results are in agreement with theoretical expectations (Figure~\ref{fig:viscosity}) and confirm the importance of this approach in the linear MPM scheme adopted here.}

    \begin{figure}
	   \centering
	   \includegraphics[width=\linewidth]{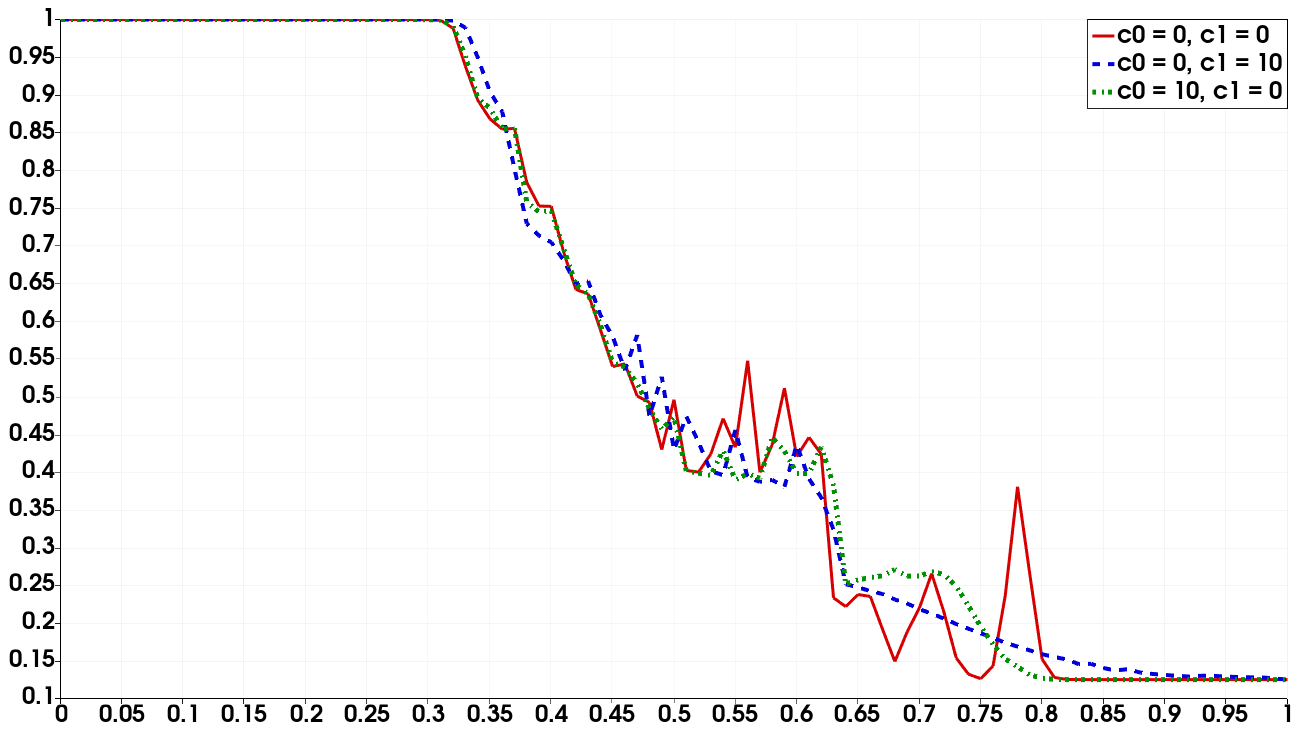}
	   \caption{Sod Shock Tube - density trend at time $t=0.143$, varying artificial viscosity coefficients.}\label{fig:viscosity}
    \end{figure}
 
	\paragraph{Resolution}
	Having addressed the applicability of the method to high-speed transonic and supersonic compressible gas dynamics, we focus on the cylinder test case and test the dependence of the results on key numerical parameters. Specifically, we vary the number of grid cells, $N_C$, and the average number of particles per cell, $ppc$, and evaluate the effect on the pressure of the flow.\\
	First, sufficiently high spatial resolution must be used to actually show shock waves; indeed, notwithstanding the use of artificial viscosity as shock capturing technique, with $N_C\sim10^3$ cells we are not able to see any shock. However, a run using $N_C\simeq10^4$ cells is able to show flow features which are in agreement with what expected from the literature, as can be seen in Figure~\ref{fig:c0_1sec}.\\
	Besides grid resolution, supersonic compressible fluid dynamics requires further care in setting the average number of particles per cell. Indeed, $ppc$ affects stability in a twofold fashion. \\
	First, for low $ppc$ values (e.g.,1), widespread oscillations are observed all along the simulation, due to almost randomly distributed empty cells. In a recent study, a $ppc$ value of 4 was found to be optimal for convergence of gas dynamics tests at lower speed~\cite{ChenTD}, and, indeed, a $ppc=4$ solves this issue. However, for supersonic flow past a static obstacle, we observed that using $ppc=4$ might cause the formation of domain regions of empty or quasi-empty cells, which gets evident as grid resolution increases, during the initial transient phase from uniform, undisturbed conditions, as can be seen from Figure~\ref{fig:dynamics}. Such emptying causes later formation of non-physical cavitation-like effects during the transient that precedes equilibrium, which alter the dynamics, and, if particularly strong, can cause also spurious shock waves, that interact with the physical ones. These events can be prevented by increasing the average particles per cell number: at Mach 3, spurious shock waves become less relevant with $ppc = 9$, and undetectable with $ppc = 16$.\\
	These effects, and in particular strong disruptions, are of no interest here, since the initial transient phase is non-physical, and derives only from the choice to start the simulations with undisturbed inflow conditions and particles position sampled from a uniform random distribution instead that from a previously computed equilibrium condition, e.g. the result of an already converged simulation. However, it could be a relevant aspect if fast changing dynamic situations have to be simulated, for example with a moving cylinder in a FSI framework, since they would alter its dynamics. In such cases, a sufficient number of particles per cell should be used.\\
	{By contrast, when interested in main equilibrium conditions, a $ppc = 4$ value appears to provide sufficient solution quality. Numerical experiments show that equilibrium is reached notwithstanding spurious instabilities in the initial transient phase, as shown in temporal averaged pressure in Figure~\ref{fig:resolution}. However, as shown in the same figure, a higher resolution allows the accurate representation of small areas, such as the ones in the immediate rear of the cylinder, where quantities, such as the average Mach number, change in a non-negligible way.}

 \begin{figure}
	\centering
	\includegraphics[width=\linewidth]{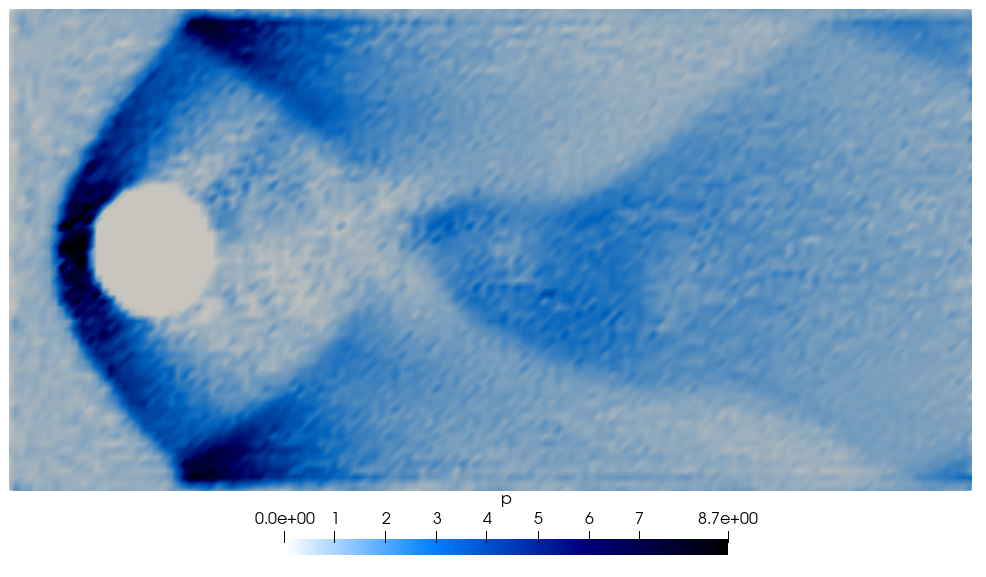}
	\caption{Mach 3 flow past a cylinder test case with $N_C = 10^4$, $ppc = 4$, pressure field at $t\simeq1s$.}\label{fig:c0_1sec}
\end{figure}
\begin{figure*}
	\centering
	\includegraphics[width=\linewidth]{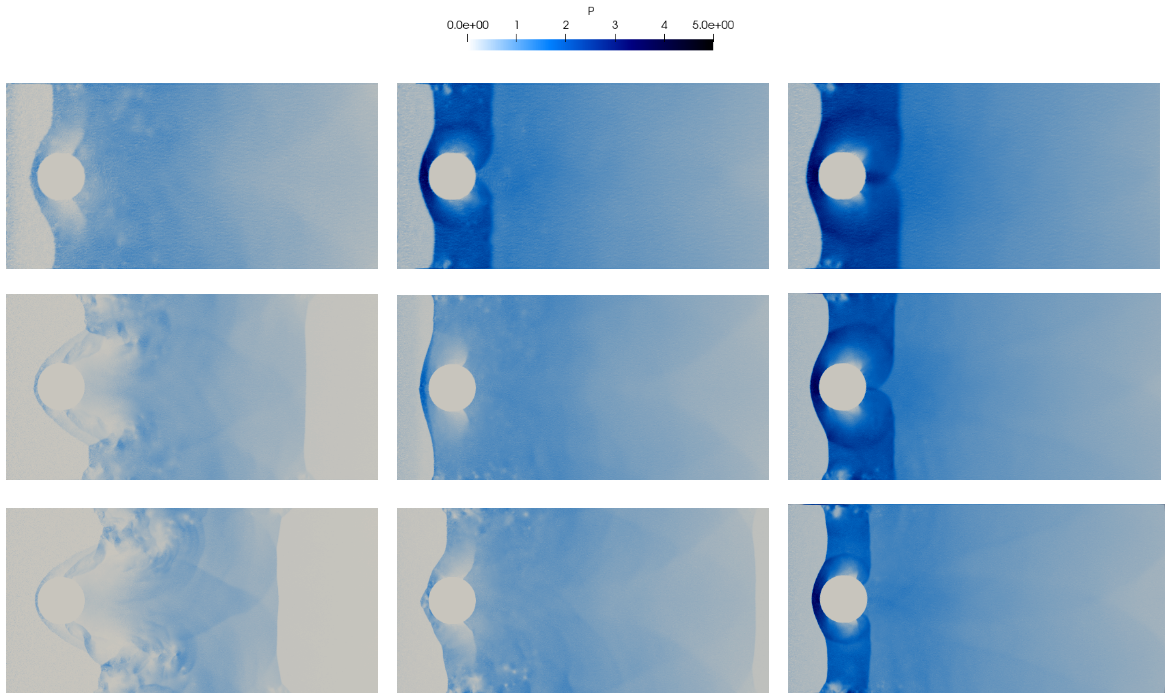}
	\caption{Mach 3 flow past a cylinder test case, various settings: from the first to the third row, $N_C$ changes from $10^5$ to $5\cdot 10^5$ and $10^6$; from the first to the third column, $ppc$ changes from 4, to 9 and to 16.}\label{fig:dynamics}
\end{figure*}
\begin{figure*}
	\centering
	\includegraphics[width=\linewidth]{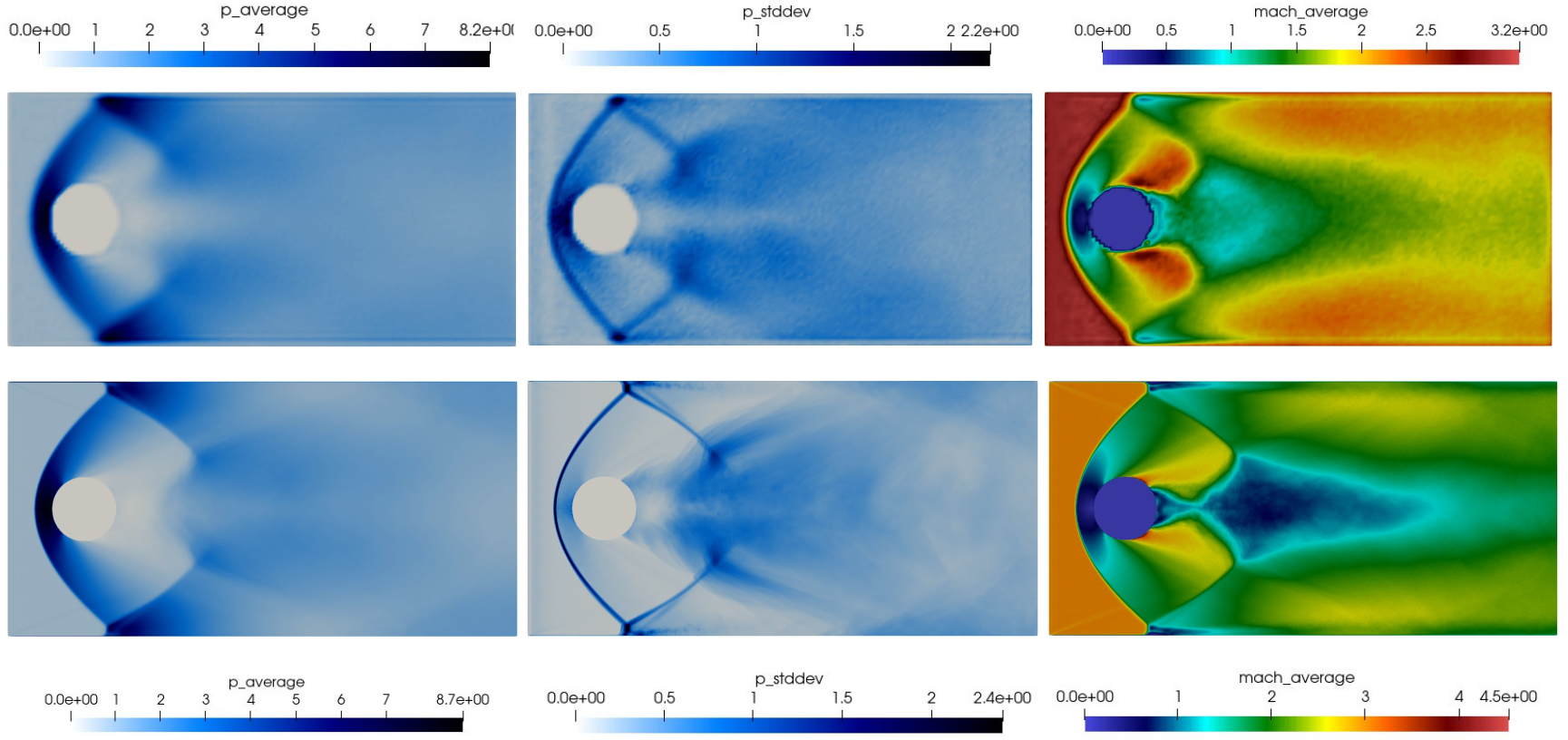}
	\caption{Mach 3 flow past a cylinder test case, at low resolution (top row, $N_C = 10^4$, $ppc = 4$) and high resolution (bottom row, $N_C = 10^6$, $ppc = 16$), from left to right: average pressure, pressure standard deviation (represented only to provide a clearer image of the shocks), average Mach number. Values computed as arithmetic average over 85 snapshots corresponding to physical simulation time from 15 seconds to 100 seconds.}\label{fig:resolution}
\end{figure*}
	
	\subsection{Computational performance}\label{subsec:perf}
	We then focus on the computational performance and scalability of the method, analysing the computational time required by simulations when varying the grid resolution and $ppc$ values. It is worth noting that the implemented parallel kernels can be particle-based or grid-based, and, respectively, assign one particle or one grid node to one GPU thread. Moreover, we implemented both mapping kernels (P2G and G2P) as particle-based kernels, since this proved to be the most efficient option.  
	
	\noindent First, a detailed profiling of each step of algorithm~\ref{algo1} reveals that the G2P and the P2G kernels take up $85\%$ of the parallel part of the code and more than $60\%$ of the whole computational time. This is due both to the relative complexity of those two kernels and to the high number of times they are called per algorithm iteration. We thus focus the scalability analysis on these kernels.\\
	The first finding is that both kernels scale linearly with the total number of particles, in agreement with their implementation (Figures~\ref{fig:profiling1_g2p} and~\ref{fig:profiling1_p2g}), while pure grid kernels, such as boundary conditions enforcement, scale with the number of nodes. 
	For this reason, we applied a global particles reordering step in the algorithm and studied its effects on performance. Indeed, the ordering is done so that the lowest global indices are assigned to the particles which are in the first grid cell after moving, followed by those in the second grid cell, and so on, moving along the grid. Thus, data which will be accessed at the next time iteration will be ordered, with contiguous particles having contiguous indexes, keeping data locality across the simulation, which is of utmost importance on our target architecture. The reordering turns out to have a significant impact on the G2P kernels, resulting in a further 4x speed-up factor, while preserving the linear scaling (Figure~\ref{fig:profiling1_g2p}).\\
	By contrast, the reordering appears to have a slightly negative effect on the P2G kernels, with slowdowns which are more evident at some particle values (Figure~\ref{fig:profiling1_p2g}). The reason can be found in the main implementation difference among the two kernels. P2G introduces a data race, that, on NVIDIA GPUs, is currently dealt with using CUDA atomicAdd, and its cost becomes more apparent with re-ordering. Indeed, even though, to the best of the authors' knowledge, NVIDIA's A100 scheduler policy is not accessible, numerical examples appear to suggest that the scheduler behaves more in a round-robin fashion rather than in a block partitioning way. So, a particle is assigned to a GPU thread, until there are some free, then a second particle is assigned, and so on. In this framework, randomisation of the global order of particles is somewhat beneficial, since it lowers the probability that particles that are in the same cells are processed simultaneously. After reordering, particles close in the domain are also close in memory, enhancing probability of a memory collision, i.e. of simultaneous access to the same data. Since in our case such access would be in read-and-write mode, it activates the atomic operation and degrades performance.\\
	Moreover, since a higher number of ordered particles per cell increases the chance of data races, the average number of particles per cell becomes an important parameter. As a matter of fact, grouping simulations results per $ppc$ value, we can confirm the trend of worsening performance at larger numbers of ordered particles in a cell. Indeed, comparing the linear fits in Figure~\ref{fig:profiling2}, we can see that in the $ppc=4$ case execution is faster than in the global fit, in the $ppc=16$ case it is slower, the $ppc=9$ being intermediate between the two. Moreover, we can see that at higher $ppc$ the computational time required by the P2G kernels deviates from the linear trend. The reason can be found in the relation between the average number of particles per cell and the number of grid nodes per cell. Indeed, as the $ppc$ value grows and the number of nodes per cell stays fixed, data locality is lost sooner in the simulation: in each cell, a higher amount of particles indexes become increasingly different from the grid indexes, leading to an increase in the beneficial effect of re-ordering on parallel performance.\\
	\\\noindent As for overall computational time, we note that its value largely depends on the requested resolution. On a NVIDIA A100 GPU, to simulate one second of gas flow around the cylinder with $4\cdot10^4$ particles requires about one second of computational time on the device, with $16\cdot10^6$ particles it requires about $3$ minutes. This is reasonably fast, especially considering that the latter resolution is rather excessive for the 2D cylinder test case. For comparison, 3D snow avalanches were recently simulated via MPM with $10$ to $30$ million particles, see~\cite{Gaume2018};  further considerations on extension to 3D are presented in section~\ref{sec:concl}.\\
	From a previous study~\cite{particles}, we expect that even better performance is achievable. However, with this work we are mostly interested in investigating the potential of MPM when simulating compressible fluids on various GPU accelerated architecture, we thus prioritise portability over efficiency, to be able to compare exactly the same code on different architecture. Portability has been reached also by means of a higher-level approach with respect to low-level CUDA-C, which introduces some level of indirection, with libraries as Thrust that rely on managed memory.\\
	In the next paragraph, we exploit the flexibility granted from this approach to test the code on different hardware architectures, both on CPUs and GPUs, and provide an evaluation of the achieved speed-up, with respect to serial and CPU-parallel versions.

     \begin{figure}
    	\centering
    	\includegraphics[width=\linewidth]{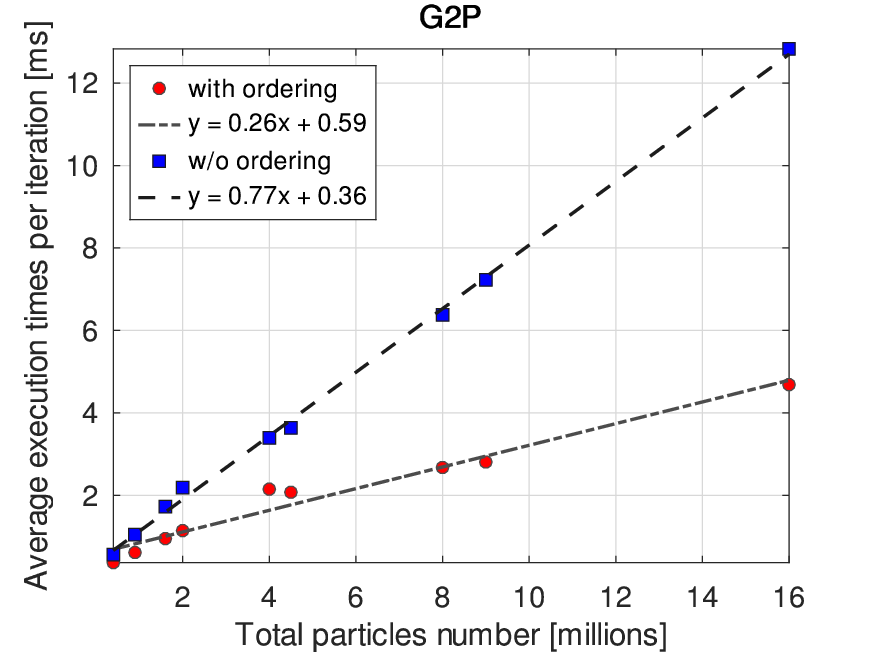}
    	\caption{Profiling of main computational kernels: G2P kernels execution time is compared for the cylinder test case, varying $N_C$ and $ppc$, considering also re-ordering effects.}\label{fig:profiling1_g2p}
    \end{figure}
    \begin{figure}
    	\centering
    	\includegraphics[width=\linewidth]{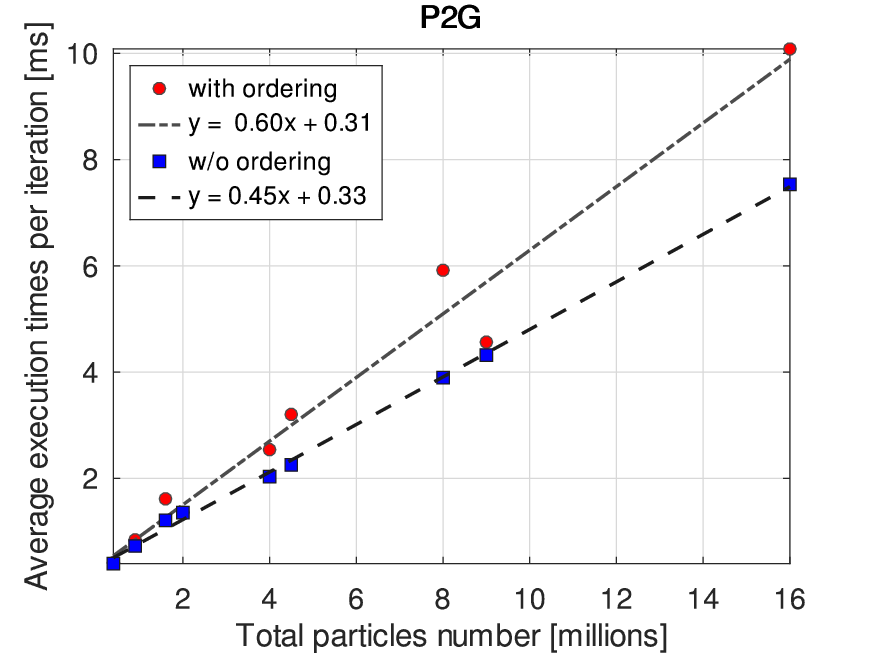}
    	\caption{Profiling of main computational kernels: P2G kernels execution time is compared for the cylinder test case, varying $N_C$ and $ppc$, considering also re-ordering effects.}\label{fig:profiling1_p2g}
    \end{figure}
    \begin{figure}
    	\centering
    	\includegraphics[width=\linewidth]{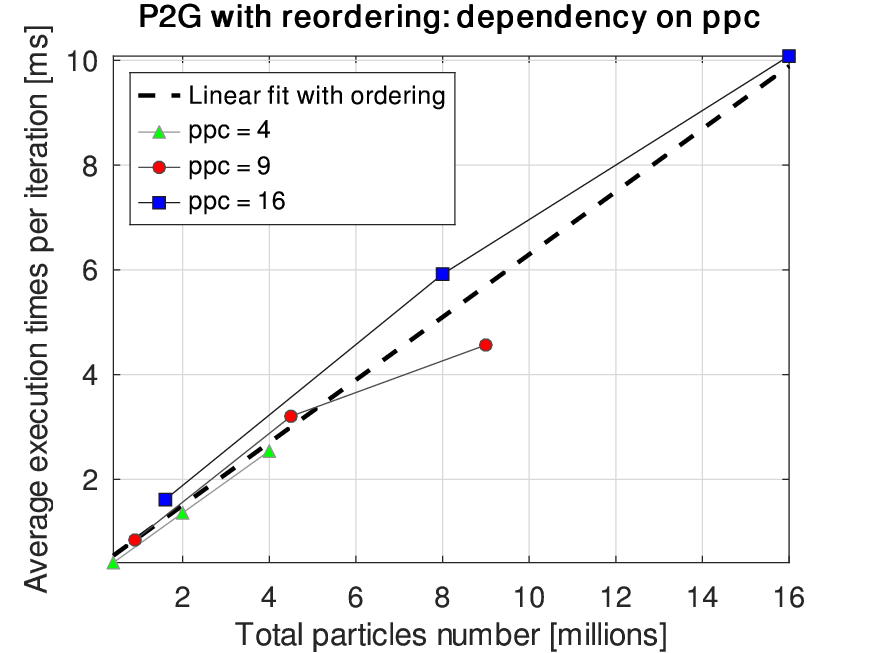}
    	\caption{Profiling of P2G kernels: effect of average particles per cell number on version with re-ordering.}\label{fig:profiling2}
    \end{figure}

	{
		\subsection{Performance portability}\label{subsec:portab}
		We test the application of the developed implementation on different hardware architectures and compare the performance, addressing the portability of the MPM code. Specifically, we execute the cylinder test case at several resolutions, on different Intel and AMD CPUs and NVIDIA and AMD GPUs, with the same setting described in the previous sections, without enforcing particles re-ordering.\\
		
		\noindent For the first test, we vary the number of cells and particles per cell and study how this affects the main computational kernels on an AMD CPU, employing $2x24$ cores in parallel (Figure~\ref{fig:cpu}).\\
		While portability is automatically granted, CPU performance turns out to be worse than single-GPU performance, even though $2x24$ CPU cores are employed in parallel, as can be seen comparing total wall-clock times in Figure~\ref{fig:speedupandscaling}; this is consistent with expectations that MPM is better suited to GPU architectures. It is anyway positive that the computational time stays linear with respect to the number of particles.\\
		As for the computational kernels:
		\begin{itemize}
			\item P2G kernels scale almost linearly with the number of total particles, in agreement with the beneficial effect of randomisation in reducing the actual execution of atomic operations;
			\item G2P kernels maintain quasi-linear scaling only if the average particles-per-cell number is kept fixed;
			\item the execution time required by all the other kernels considered together is considerably smaller than the time required by P2G and G2P ones. This confirms the  computational relevance of P2G and G2P also on different hardware.
		\end{itemize}
		The main difference with respect to the GPU test lies in the comparison between P2G and G2P performance, with the former significantly slower than the latter on the CPU. This is most probably due to our choice of writing an ad-hoc {\ttfamily atomicAdd} function, based on {\ttfamily std::atomic\_ref}, with the same signature of the more performing CUDA's {\ttfamily atomicAdd}, to enhance portability and guarantee parallelism. An algorithm to directly avoid the data-race present in P2G-like kernels, and, thus, to provide a natively portable, parallel and atomic operation free implementation, is proposed in the next section.\\

        \begin{figure*}
        	\centering
        	\includegraphics[width=\linewidth]{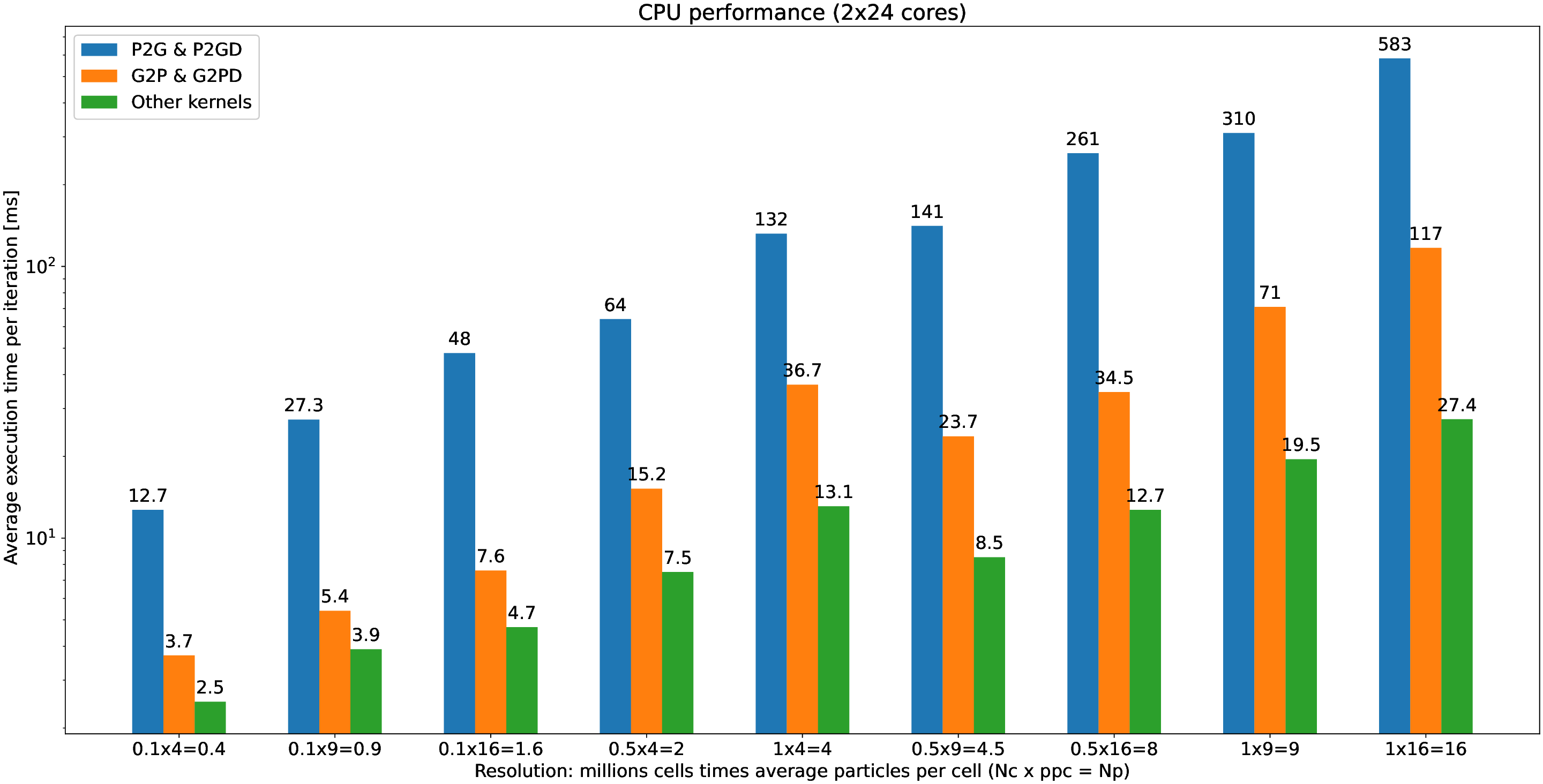}
        	\caption{Profiling of main computational kernels: total G2P and P2G kernels execution time is compared for the cylinder test case, varying $N_C$ and $ppc$, without re-ordering, on $2x24$ CPU cores (AMD 7402 CPU).}\label{fig:cpu}
        \end{figure*}
        \begin{figure}
        	\centering
        	\includegraphics[width=\linewidth]{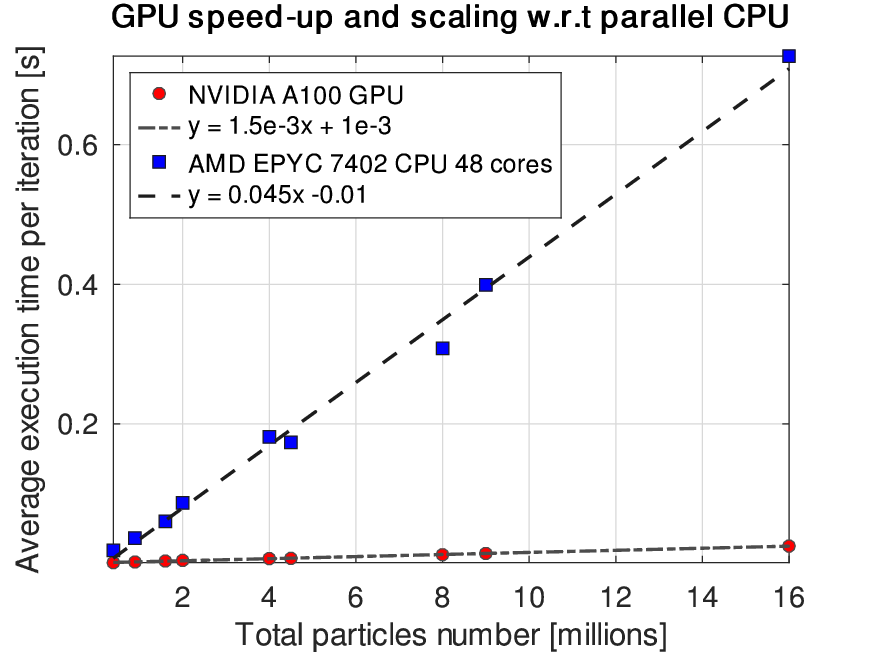}
        	\caption{Computational time scaling and speed-up varying number of total particles and parallel architecture.}\label{fig:speedupandscaling}
        \end{figure}

		\noindent For the second test, we focus on the heaviest test case (the cylinder with $16$ million particles) and simulate one physical second on different architectures, comparing the total wall-clock time, referred to the time required by the device, being either a CPU or a GPU; see Figure~\ref{fig:3_hist}. It is interesting to note that the chosen implementation approach allows the porting of the code to AMD GPUs without any modification.\\

        \begin{figure*}
        	\centering
        	\includegraphics[width=\linewidth]{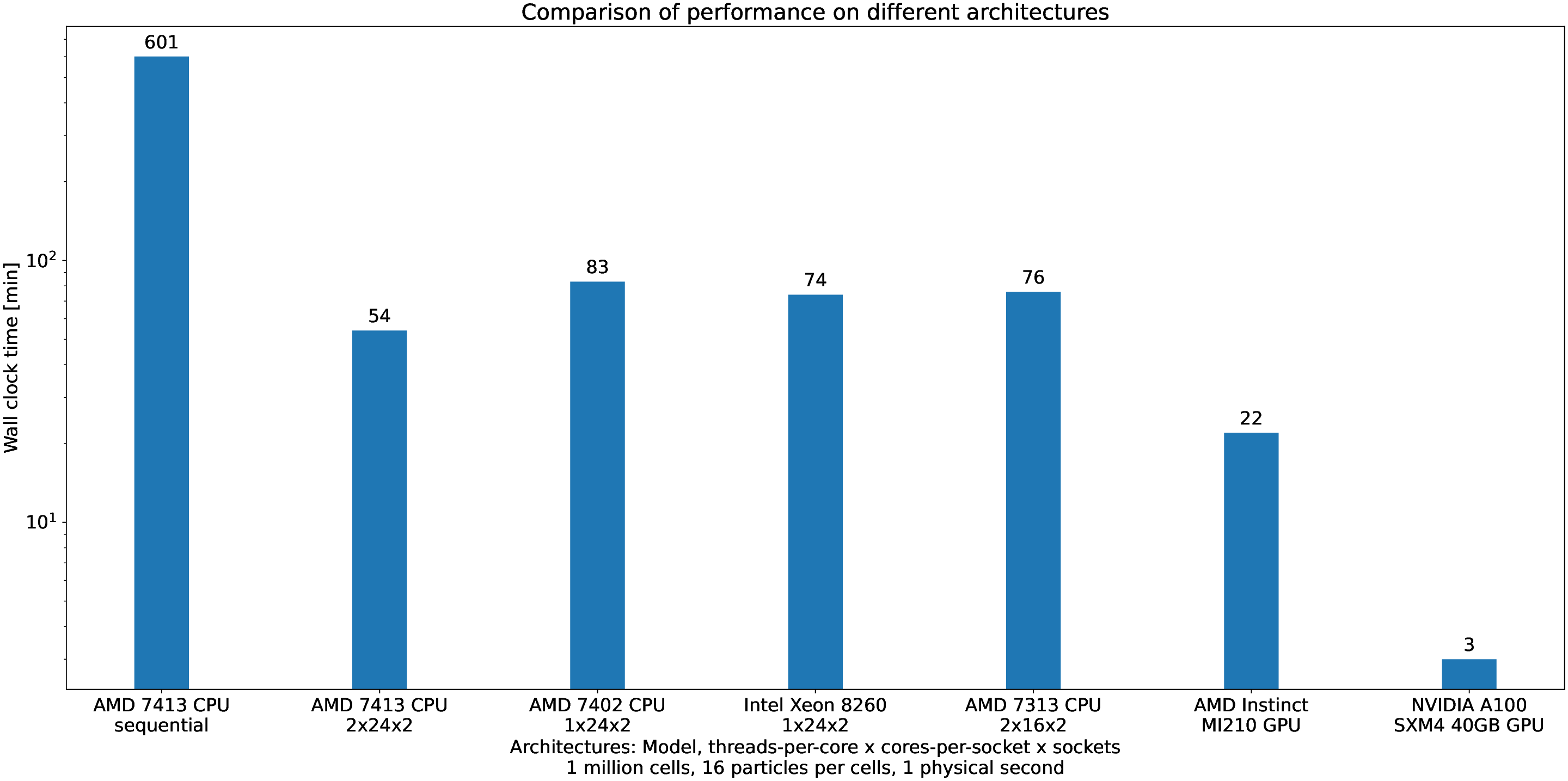}
        	\caption{Comparison of performance on different architectures: overall computational time to simulate 1 physical second of the cylinder test case with $N_C = 10^6$, $ppc = 16$.}\label{fig:3_hist}
        \end{figure*}
  
		\noindent Then, we focus on the comparison between the two GPU architectures.\\
		We can notice that at low computational load, performance is basically equivalent, especially if the higher degree of code optimisation allowed by the NVIDIA compilers is taken into account, even though the AMD GPU seems to be more sensible to the kernels with atomicAdd; see Figure~\ref{fig:4a_detailed_hist}.\\
		Instead, at heavier loads, despite the comparable specs, the AMD MI210 starts to display longer wall-clock times than the NVIDIA A100, see Figure~\ref{fig:4b_hist} for a comparison of the overall computational time on the device. This difference might be due to different aspects, such as having developed the code, and taken performance based design choices, only on NVIDIA A100 GPUs, and having ported it to different hardware afterwards.
		However, in the literature, odd performance issues have been found in high-load scenarios on AMD MI100 GPUs that are currently being investigated from AMD itself~\cite{rocstdpararxiv}, so we cannot exclude to be in a similar situation here.\\
			In any case, GPU performance is overall competitive with respect to CPU. For example, the heaviest test case is 7 to 8 times slower on MI210 than on A100, but anyway 3 to 4 times faster than parallel execution via Intel OneAPI-TBB on Intel Xeon with 2x24 cores.\\

		\noindent Lastly, it is interesting to note that, with the implemented approach, the code is portable also to non-HPC architectures, as common commercial processors, and that the wall-clock time is not excessive as long as the computational load is low enough, as can be seen in Figure~\ref{fig:Figure_1}.}\\
		Most importantly, these results enable us to confirm the portability of the approach.

        \begin{figure}
        	\centering
        	\includegraphics[width=\linewidth]{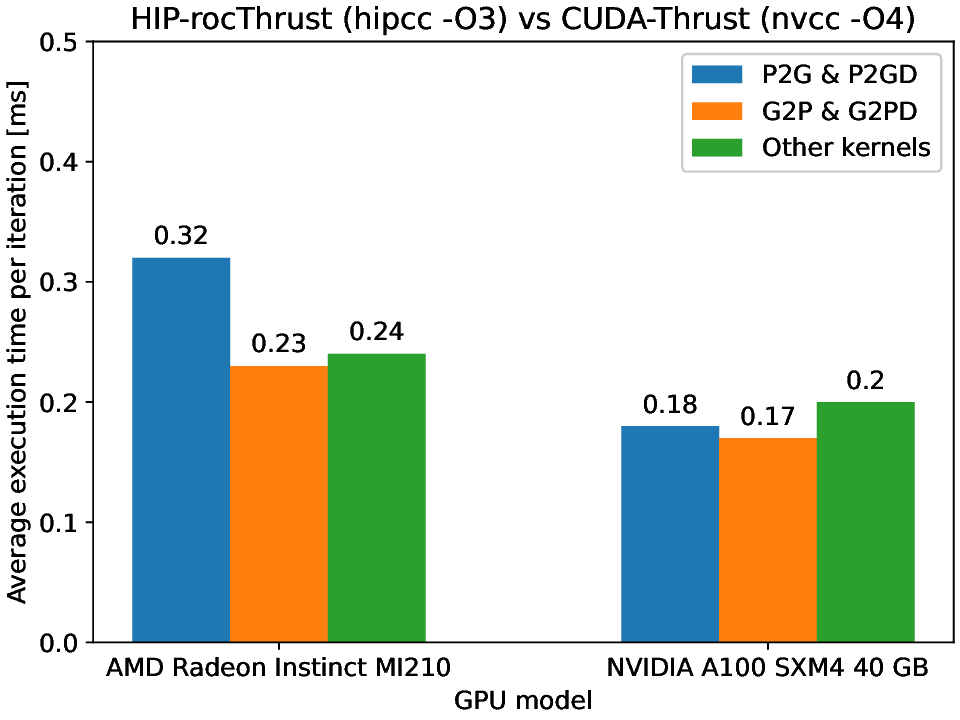}
        	\caption{Comparison of performance on different GPUs: analysis of the total wall-clock computational time for a low resolution test case $N_C = 10^4$, $ppc = 4$, corresponding to less then one computational second to simulate 1 physical second.}\label{fig:4a_detailed_hist}
        \end{figure}
        \begin{figure*}
        	\centering
        	\includegraphics[width=\linewidth]{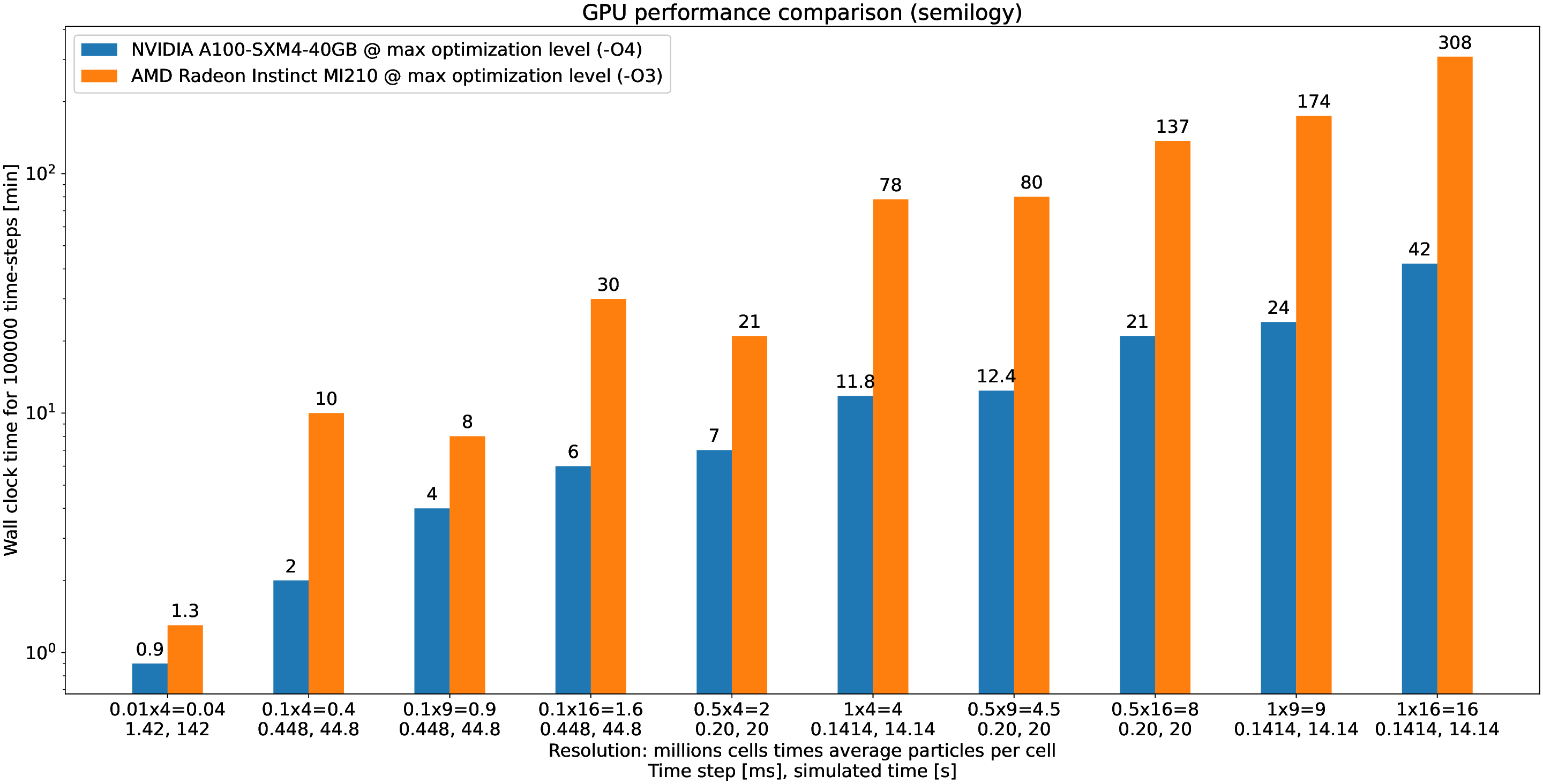}
        	\caption{Comparison of performance on different GPUs: overall computational time required for $Nt=10^5$ time loop iterations, varying resolution from $N_C = 10^4$, $ppc = 4$ to $N_C = 10^6$, $ppc = 16$.}\label{fig:4b_hist}
        \end{figure*}
        \begin{figure}
        	\centering
        	\includegraphics[width=\linewidth]{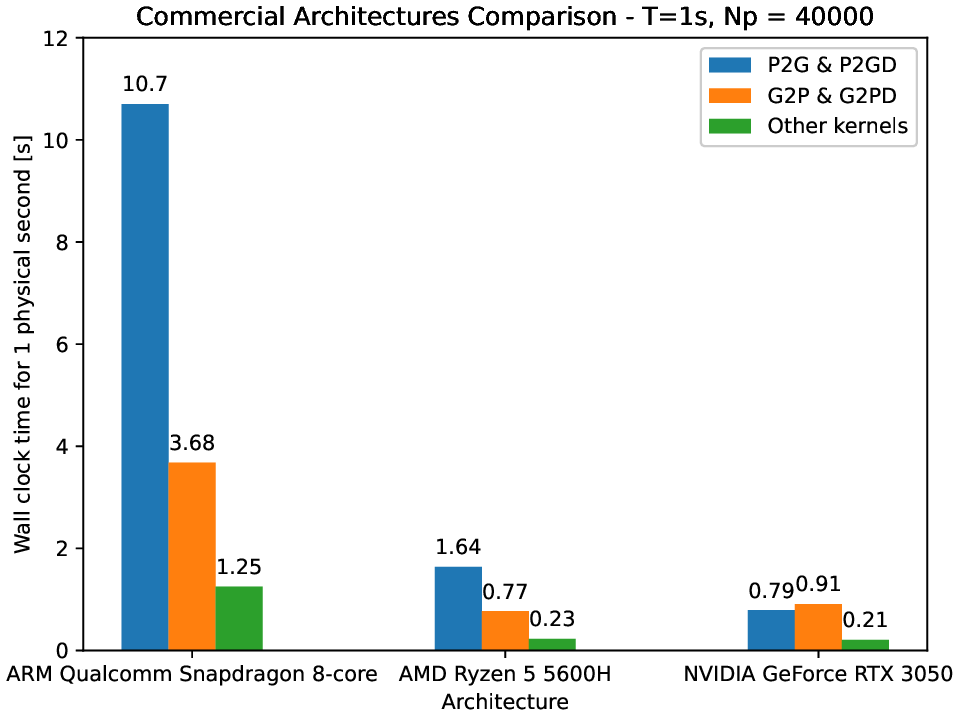}
        	\caption{Comparison of performance on different commercial, non HPC architectures: analysis of the total wall-clock computational time for a low resolution test case $N_C = 10^4$, $ppc = 4$.}\label{fig:Figure_1}
        \end{figure}
 
	\section{Discussion and conclusion}\label{sec:concl}
	\noindent This paper has showcased the potential of a massively parallel, portable implementation of the Material Point Method for compressible flows. Insight has been given on algorithm adaptations which will provide further benefits, moreover, the simulation of standard test cases of compressible flow has provided results comparable with the literature and with established software runs with the same polynomial order ($\mathbb{Q}_1$ FEM basis functions on Cartesian tensor product grids) and number of degrees of freedom. These results show that MPM is able to simulate compressible gas dynamics, giving good results with a prototype which has been developed stemming from existing implementations~\cite{York,Sandia}, and serve as further step towards the design of a highly efficient MPM-based transonic and supersonic FSI solver. \\
	As known from the literature~\cite{YorkThesis,Steffen} and as also noted in a previous implementation~\cite{particles}, $\mathbb Q 1$ MPM suffers from strong numerical oscillation issues, especially in the vicinity of shocks. The issue is tackled in this paper by adding numerical viscosity in the form of additional terms in the pressure field~\cite{MPMBook}. A more natural and accurate choice could be obtained by using quadratic B-Spline basis functions~\cite{Tielen,AdVreview}, a code feature which is currently under development\footnote{\href{https://github.com/carlodefalco/quadgrid}{https://github.com/carlodefalco/quadgrid}}.\\
	Performance tests on NVIDIA A100 GPUs have confirmed that the main computational steps of the MPM algorithm are the Particles-to-Grid and Grid-to-Particles kernels, and that they scale well with the size of the problem. Moreover, adding a reordering step is beneficial to data locality and, consequently, to the performance, while data races in the P2G kernels can be tackled CUDA's atomics at acceptable cost. Finally, it has been shown that an alternative to atomics exists, and that it could be explored in the look for a higher degree of performance portability, especially on CPU architectures. Ultimately,  this approach has shown to be easily portable to different hardware architectures, as different vendors GPUs and CPUs.\\
	
	\noindent Several future developments are planned in connection with the implementation and modelling choices made in this paper:
	\begin{itemize}
		\item On the modelling side, more complex test cases could be simulated, eventually leading to a 3D extension and Fluid-Structure Interaction models. In addition, the effect on parallel performance of higher order B-Splines, currently under development, will need to be carefully investigated.{\\
			A 3D extension will facilitate a further growth in computational load, but this will not introduce conceptual main changes in the code, thanks also to the high level, portable approach, which does not refer explicitly to specific, small size memories. Indeed, extension to the 3D case would require only slight modifications of the main algorithm, which would consist in repeating section~\ref{sec:alg_and_impl} steps also in the z direction where necessary and in adding the relative variables to the {\ttfamily mpm} struct and the z component to the P2GD and G2PD kernels. In particular, the computational load of the 2D 16 million particles cylinder simulation of section~\ref{sec:res} is exemplary of the one of 3D realistic test cases. Considering the extra memory in the z-direction, our 2D simulation performance is representative of the one of a 3D simulation with at least 10 million particles; a resolution which was used for recent, realistic scenarios MPM simulations in the literature~\cite{Gaume2018}. As a matter of fact, adding one spatial dimension would not have a large impact on the code, since the largest share of the computational time is taken by kernels which we implement as Lagrangian particles-index based ones. \\
				However, as data files become larger, a 3D extension }could justify a more efficient I/O, e.g., moving from JSON to HDF5. Simulations characterised by a higher computational load would also further benefit from a future multi-GPUs approach, that, to be optimal, would require the introduction of a load balance mechanism among the GPUs~\cite{Ruggirello2014}.\\
			The upgrade to a higher order MPM via proper shape function, as B-Splines, would not require a significant change in the numerical method, but will introduce some challenges linked to parallelism.\\
			In particular, 
			expansion of shape function support to neighbouring grid elements will introduce halo overlapping, leading to an increase in communication costs, as well as to a constraint on the maximum level of achievable parallel granularity.\\
			Few works on this topic exist, such as~\cite{MultiGPU}, and, since they simulate continuum mechanics materials instead of compressible gases, they can benefit from sparse data structures. However, it is already possible to anticipate that shape function regularity in the current MPM formulation could grow efficiently only up to a certain maximum, such as quadratic or cubic B-Spline, and start to have an impact on performance as shape function support expands.
		\item For the algorithm implementation, the use of atomic operations to handle data races has been chosen for simplicity, {also because it has shown an acceptable impact on computational time and scalability on NVIDIA A100 GPUs and a simple portability to other architectures (section~\ref{sec:res})}. To reach the full parallelisation capabilities proved by the G2P kernels also for the P2G ones, P2G algorithm adaptations will be necessary. An interesting and and more portable route would involve avoiding atomics, and writing a \textit{grid colouring} algorithm that can be implemented in an architecture-independent way, relying only on standard libraries functions. Its main steps are:
		\begin{enumerate}
			\item Build grid-to-particle connectivity
			\begin{enumerate}[(i)]
				\item Sort particle IDs by cell index
				\item Count particles per cell
				\item Perform a cumulative sum of particle counts
			\end{enumerate}
			\item Assign a colour to each cell, in such a way that neighbouring cells have different colours; then, for each  colour:
			\begin{enumerate}
				\item Perform a {\ttfamily for\_each} on cells
				\item Call kernels~\eqref{eq:p2g_gen} or~\eqref{eq:p2gd_gen}
			\end{enumerate}
		\end{enumerate}
		This way we would iterate on the vector holding the list of particles in each cell, and then we would perform only a small amount of serial operations inside each cell, without any data race.
		\item On the High Performance Computing side, avenues of development include the support of multi-GPU and multi-node architectures via MPI, and a grid-particles adaptive refinement through a quad-tree approach~\cite{adapt1,adapt2piufract}. Moreover, currently the code employs Structure-of-Arrays for the memory, since, differently from arrays of arbitrary structures, it provides coalescing, which improves memory efficiency~\cite{GTC}. However, recent developments on GPUs computing literature~\cite{MultiGPU,RealTime} show that the best performance is achieved by adopting an Array of SoA (AoSoA). This possibility could be explored when addressing the MPI-based multi-GPU implementation. Indeed, the current data structure gives a conceptually straightforward extension towards multi-GPU implementations, where a further level of parallelisation through an MPI-based domain decomposition will make each GPU hold a SoA as the current one, globally giving rise to an AoSoA data structure. 
	\end{itemize}

	\FloatBarrier
	\section*{Acknowledgements}
Research was sponsored by the Leonardo Labs, Leonardo S.p.A., through the PhD scholarship ``Multi-GPUs based PIC / MPM methods for Compressible Flows" funded at Politecnico di Milano and by ICSC - Centro Nazionale di Ricerca in High Performance Computing, Big Data, and Quantum Computing funded by European Union - NextGenerationEU.
		{The simulations discussed in this work were, in part, performed on the HPC Cluster of the Department of Mathematics of Politecnico di Milano which was funded by MUR grant Dipartimento di Eccellenza 2023-2027. C. de Falco and P. J. Baioni are members of the GNCS group of INdAM.}
		Views and conclusions contained in this document are those of the author and should not be interpreted as representing the official policies, either expressed or implied, of Leonardo Labs, ICSC, {MUR or GNCS-INdAM}.  
	%
	\section*{Conflict of interest}
	
	The authors declare that they have no conflict of interest.
	
	
	\bibliographystyle{spmpsci}      
	\bibliography{Portable_Parallel_MPM_Compressible_Flows}   

\begin{thebibliography}{10}
\providecommand{\url}[1]{{#1}}
\providecommand{\urlprefix}{URL }
\expandafter\ifx\csname urlstyle\endcsname\relax
  \providecommand{\doi}[1]{DOI~\discretionary{}{}{}#1}\else
  \providecommand{\doi}{DOI~\discretionary{}{}{}\begingroup
  \urlstyle{rm}\Url}\fi

\bibitem{LargeDeform1}
Andersen, S., Andersen, L.: Modelling of landslides with the material-point
  method.
\newblock Computational Geosciences \textbf{14}(1), 137--147 (2010)

\bibitem{particles}
Baioni, P.J., Benacchio, T., L.Capone, de~Falco, C.: Gpus based material point
  method for compressible flows.
\newblock VIII International Conference on Particle-Based Methods  (2023)

\bibitem{normali}
Bardenhagen, S., Brackbill, J., Sulsky, D.: The material-point method for
  granular materials.
\newblock Computer Methods in Applied Mechanics and Engineering
  \textbf{187}(3), 529--541 (2000)

\bibitem{GIMP}
Bardenhagen, S.G., Kober, E.M.: The generalized interpolation material point
  method.
\newblock Computer Modeling in Engineering \& Sciences \textbf{5}(6), 477--496
  (2004)

\bibitem{Berzins}
Berzins, M.: Energy conservation and accuracy of some mpm formulations.
\newblock Computational Particle Mechanics  (2022)

\bibitem{gpgpu2003}
Bolz, J., Farmer, I., Grinspun, E., Schr\"{o}der, P.: Sparse matrix solvers on
  the gpu: conjugate gradients and multigrid.
\newblock ACM Trans. Graph. \textbf{22}(3), 917–924 (2003).
\newblock \doi{10.1145/882262.882364}

\bibitem{FLIP}
Brackbill, J., Ruppel, H.: Flip: A method for adaptively zoned,
  particle-in-cell calculations of fluid flows in two dimensions.
\newblock Journal of Computational Physics \textbf{65}(2), 314--343 (1986)

\bibitem{GPUcfr}
Breyer, M., Van~Craen, A., Pfl\"{u}ger, D.: A comparison of sycl, opencl, cuda,
  and openmp for massively parallel support vector machine classification on
  multi-vendor hardware.
\newblock In: International Workshop on OpenCL, IWOCL'22. Association for
  Computing Machinery, New York, NY, USA (2022)

\bibitem{sdf}
Bruchon, J., Digonnet, H., Coupez, T.: Using a signed distance function for the
  simulation of metal forming processes: Formulation of the contact condition
  and mesh adaptation. from a lagrangian approach to an eulerian approach.
\newblock International Journal for Numerical Methods in Engineering
  \textbf{78}(8), 980--1008 (2009)

\bibitem{AdV2024}
Buckland, E., Nguyen, V.P., de~Vaucorbeil, A.: Easily porting material point
  methods codes to gpu.
\newblock Computational Particle Mechanics  (2024).
\newblock \doi{10.1007/s40571-024-00768-1}.
\newblock \urlprefix\url{https://doi.org/10.1007/s40571-024-00768-1}

\bibitem{gas}
Cao, Y., Chen, Y., Li, M., Yang, Y., Zhang, X., Aanjaneya, M., Jiang, C.: An
  efficient b-spline lagrangian/eulerian method for compressible flow, shock
  waves, and fracturing solids.
\newblock ACM Trans. Graph. \textbf{41}(5) (2022)

\bibitem{sdfmox}
Carrara, D., Regazzoni, F., Pagani, S.: Implicit neural field reconstruction on
  complex shapes from scattered and noisy data.
\newblock MOX Technical Report  (2024).
\newblock
  \urlprefix\url{https://mox.polimi.it/reports-and-books/publication-results/?id=1250}

\bibitem{Sandia}
Chen, Z., Brannon, R.M.: An evaluation of the material point method.
\newblock Sandia National Lab. Technical Report  (2002).
\newblock \urlprefix\url{https://www.osti.gov/biblio/793336}

\bibitem{adapt2piufract}
Cheon, Y.J., Kim, H.G.: An adaptive material point method coupled with a
  phase-field fracture model for brittle materials.
\newblock International Journal for Numerical Methods in Engineering
  \textbf{120}(8), 987--1010 (2019)

\bibitem{CLAYTON2023111926}
Clayton, B., Guermond, J.L., Maier, M., Popov, B., Tovar, E.J.: Robust
  second-order approximation of the compressible euler equations with an
  arbitrary equation of state.
\newblock Journal of Computational Physics \textbf{478}, 111926 (2023)

\bibitem{Coombs}
Coombs, W.M., Charlton, T.J., Cortis, M., Augarde, C.E.: Overcoming volumetric
  locking in material point methods.
\newblock Computer Methods in Applied Mechanics and Engineering \textbf{333},
  1--21 (2018)

\bibitem{BrezziFortin}
Daniele~Boffi Franco~Brezzi, M.F.: Mixed Finite Element Methods and
  Applications.
\newblock Springer Berlin, Heidelberg (2013).
\newblock \doi{https://doi.org/10.1007/978-3-642-36519-5}

\bibitem{TLMPM}
{de Vaucorbeil}, A., Nguyen, V.P., Hutchinson, C.R.: A total-lagrangian
  material point method for solid mechanics problems involving large
  deformations.
\newblock Computer Methods in Applied Mechanics and Engineering \textbf{360},
  112783 (2020)

\bibitem{AdVreview}
{de Vaucorbeil}, A., Nguyen, V.P., Sinaie, S., Wu, J.Y.: Chapter two - material
  point method after 25 years: Theory, implementation, and applications.
\newblock In: S.P. Bordas, D.S. Balint (eds.) Advances in Applied Mechanics,
  vol.~53, pp. 185--398. Elsevier (2020).
\newblock \doi{https://doi.org/10.1016/bs.aams.2019.11.001}

\bibitem{vandyke}
Dyke, M.V.: An Album of Fluid Motion.
\newblock ParabolicPress, Inc. (2008)

\bibitem{nomoreporting}
Elafrou, A., Larkin, J.: No More Porting: GPU Computing with Standard C++ and
  Fortran.
\newblock NVIDIA - GTC Spring 2023 (2023)

\bibitem{RealTime}
Fei, Y., Huang, Y., Gao, M.: Principles towards real-time simulation of
  material point method on modern gpus.
\newblock CoRR \textbf{abs/2111.00699} (2021)

\bibitem{PolyPIC}
Fu, C., Guo, Q., Gast, T., Jiang, C., Teran, J.: A polynomial particle-in-cell
  method.
\newblock ACM Trans. Graph. \textbf{36}(6) (2017)

\bibitem{GPU}
Gao, M., Wang, X., Wu, K., Pradhana, A., Sifakis, E., Yuksel, C., Jiang, C.:
  Gpu optimization of material point methods.
\newblock ACM Trans. Graph. \textbf{37}(6) (2018)

\bibitem{Gaume2018}
Gaume, J., Gast, T., Teran, J., van Herwijnen, A., Jiang, C.: Dynamic anticrack
  propagation in snow.
\newblock Nature Communications \textbf{9}(1), 3047 (2018).
\newblock \doi{10.1038/s41467-018-05181-w}.
\newblock \urlprefix\url{https://doi.org/10.1038/s41467-018-05181-w}

\bibitem{Goktekin}
Goktekin, T.G., Bargteil, A.W., O'Brien, J.F.: A method for animating
  viscoelastic fluids.
\newblock ACM Trans. Graph. \textbf{23}(3), 463–468 (2004)

\bibitem{MaierLike}
Guermond, J.L., Nazarov, M., Popov, B., Tomas, I.: Second-order invariant
  domain preserving approximation of the euler equations using convex limiting.
\newblock SIAM Journal on Scientific Computing \textbf{40}(5), A3211--A3239
  (2018)

\bibitem{Hariri}
Hariri, F., Tran, T., Jocksch, A., Lanti, E., Progsch, J., Messmer, P.,
  Brunner, S., Gheller, C., Villard, L.: A portable platform for accelerated
  pic codes and its application to gpus using openacc.
\newblock Computer Physics Communications \textbf{207}, 69--82 (2016)

\bibitem{Harlow}
Harlow Francis H.;~Evans, M.W., Harris David~E., J.: The particle-in-cell
  method for two-dimensional hydrodynamic problems.
\newblock Report LAMS-2082 of the Los Alamos Scientifc Laboratory  (1956)

\bibitem{GTC}
Hoberock, J.: Thrust by example: Advanced features and techniques.
\newblock In: GPU Technology Conference. NVIDIA (2010)

\bibitem{MLSMPM}
Hu, Y., Fang, Y., Ge, Z., Qu, Z., Zhu, Y., Pradhana, A., Jiang, C.: A moving
  least squares material point method with displacement discontinuity and
  two-way rigid body coupling.
\newblock ACM Transactions on Graphics \textbf{37}(4), 150 (2018)

\bibitem{CJangthesis}
Jiang, C.: The material point method for the physics-based simulation of solids
  and fluid.
\newblock Ph.D. thesis, University of California, Los Angeles (2015)

\bibitem{APIC}
Jiang, C., Schroeder, C., Teran, J.: An angular momentum conserving
  affine-particle-in-cell method.
\newblock Journal of Computational Physics \textbf{338}, 137--164 (2017)

\bibitem{Schroeder2024}
Jiang, H., Schroeder, C.: Second order accurate particle-in-cell discretization
  of the navier-stokes equations.
\newblock Journal of Computational Physics \textbf{518}, 113302 (2024).
\newblock \doi{https://doi.org/10.1016/j.jcp.2024.113302}.
\newblock
  \urlprefix\url{https://www.sciencedirect.com/science/article/pii/S0021999124005503}

\bibitem{Ruggirello2014}
Kevin P.~Ruggirello, S.C.S.: A comparison of parallelization strategies for the
  material point method.
\newblock In: 11th World Congress on Computational Mechanics (WCCM XI) and 5th
  European Conference on Computational Mechanics (ECCM V) and 6th European
  Conference on Computational Fluid Dynamics (ECFD VI), 20-25 July 2014,
  Barcelona, Spain (2014)

\bibitem{sonnendrucker}
Kraus, M., Kormann, K., Morrison, P.J., Sonnendr{\"u}cker, E.: Gempic:
  geometric electromagnetic particle-in-cell methods.
\newblock Journal of Plasma Physics \textbf{83}(4), 905830401 (2017)

\bibitem{gpgpu2005}
Kr\"{u}ger, J., Westermann, R.: Linear algebra operators for gpu implementation
  of numerical algorithms.
\newblock ACM Trans. Graph. \textbf{22}(3), 908–916 (2003).
\newblock \doi{10.1145/882262.882363}

\bibitem{mu_art2}
Landshoff, R.: A numerical method for treating fluid flow in the presence of
  shocks.
\newblock Los Alamos National Lab. (LANL) Technical Report  (1955).
\newblock \doi{10.2172/4364774}.
\newblock \urlprefix\url{https://www.osti.gov/biblio/4364774}

\bibitem{FutureDirections}
Laskowski, G., Kopriva, J., Michelassi, V., Shankaran, S., Paliath, U.,
  Bhaskaran, R., Wang, Q., Talnikar, C., Wang, Z., Jia, F.: Future directions
  of high fidelity cfd for aerothermal turbomachinery analysis and design.
\newblock 46th AIAA Fluid Dynamics Conference  (2016)

\bibitem{rocstdpararxiv}
Lin, W.C., McIntosh-Smith, S., Deakin, T.: Preliminary report: Initial
  evaluation of stdpar implementations on amd gpus for hpc (2024).
\newblock \urlprefix\url{https://arxiv.org/abs/2401.02680}

\bibitem{ryujin}
Maier, M., Kronbichler, M.: Efficient parallel 3d computation of the
  compressible euler equations with an invariant-domain preserving second-order
  finite-element scheme.
\newblock ACM Trans. Parallel Comput. \textbf{8}(3) (2021)

\bibitem{matthias_maier_2020_3698223}
Maier, M., Tomas, I.: tamiko/step-69: step-69 v20200305 (2020).
\newblock \doi{10.5281/zenodo.3698223}.
\newblock \urlprefix\url{https://doi.org/10.5281/zenodo.3698223}

\bibitem{IGAMPM}
Moutsanidis, G., Long, C.C., Bazilevs, Y.: Iga-mpm: The isogeometric material
  point method.
\newblock Computer Methods in Applied Mechanics and Engineering \textbf{372},
  113346 (2020)

\bibitem{CUDAC}
NVIDIA: CUDA C++ Programming Guide.
\newblock https://docs.nvidia.com/cuda/cuda-c-programming-guide/index.html
  (2022)

\bibitem{A100}
NVIDIA: NVIDIA A100 Tensor Core GPU Architecture.
\newblock
  https://resources.nvidia.com/en-us-genomics-ep/ampere-architecture-white-paper?xs=169656
  (2022)

\bibitem{NvidiaNvc++}
NVIDIA: HPC Compilers - C++ parallel algorithm.
\newblock
  https://docs.nvidia.com/hpc-sdk/compilers/c++-parallel-algorithms/index.html
  (2023)

\bibitem{Thrust}
NVIDIA: Thrust Release 12.3.
\newblock https://docs.nvidia.com/ (Nov 14, 2023)

\bibitem{CPDI}
Sadeghirad, A., Brannon, R.M., Burghardt, J.: A convected particle domain
  interpolation technique to extend applicability of the material point method
  for problems involving massive deformations.
\newblock International Journal for Numerical Methods in Engineering
  \textbf{86}(12), 1435--1456 (2011)

\bibitem{quadtree}
Samet, H.: The quadtree and related hierarchical data structures.
\newblock ACM Computing Surveys \textbf{16} (1984)

\bibitem{spgrid}
Setaluri, R., Aanjaneya, M., Bauer, S., Sifakis, E.: Spgrid: a sparse paged
  grid structure applied to adaptive smoke simulation.
\newblock ACM Trans. Graph. \textbf{33}(6) (2014).
\newblock \doi{10.1145/2661229.2661269}

\bibitem{Sod}
Sod, G.A.: {A Survey of Several Finite Difference Methods for Systems of
  Nonlinear Hyperbolic Conservation Laws}.
\newblock {Journal of Computational Physics} \textbf{27}(1), 1--31 (1978)

\bibitem{SteffenThesis}
Steffen, M.: Analysis-guided improvements of the material point method  (2009).
\newblock
  \urlprefix\url{http://www.sci.utah.edu/publications/Ste2009a/Steffen_PhDThesis2009.pdf}

\bibitem{Steffen}
Steffen, M., Kirby, R.M., Berzins, M.: Analysis and reduction of quadrature
  errors in the material point method (mpm).
\newblock International Journal for Numerical Methods in Engineering
  \textbf{76}(6), 922--948 (2008)

\bibitem{StomakhinSnow}
Stomakhin, A., Schroeder, C., Chai, L., Teran, J., Selle, A.: A material point
  method for snow simulation.
\newblock ACM Trans. Graph. \textbf{32}(4) (2013).
\newblock \doi{10.1145/2461912.2461948}

\bibitem{Stomakhin}
Stomakhin, A., Schroeder, C., Jiang, C., Chai, L., Teran, J., Selle, A.:
  Augmented mpm for phase-change and varied materials.
\newblock ACM Transactions on Graphics \textbf{33}(4) (2014)

\bibitem{ChenTD}
Su, Y.C., Tao, J., Jiang, S., Chen, Z., Lu, J.M.: Study on the fully coupled
  thermodynamic fluid--structure interaction with the material point method.
\newblock Computational Particle Mechanics \textbf{7}(2), 225--240 (2020)

\bibitem{SulskyChenSchreyer1994}
Sulsky, D., Chen, Z., Schreyer, H.: A particle method for history-dependent
  materials.
\newblock Computer Methods in Applied Mechanics and Engineering
  \textbf{118}(1-2), 179--196 (1994)

\bibitem{iMPM}
Sulsky, D., Gong, M.: Improving the Material-Point Method, pp. 217--240.
\newblock Springer International Publishing, Cham (2016)

\bibitem{adapt1}
Tan, H., Nairn, J.A.: Hierarchical, adaptive, material point method for dynamic
  energy release rate calculations.
\newblock Computer Methods in Applied Mechanics and Engineering
  \textbf{191}(19), 2123--2137 (2002)

\bibitem{Tielen}
Tielen, R., Wobbes, E., Möller, M., Beuth, L.: A high order material point
  method.
\newblock Procedia Engineering \textbf{175}, 265--272 (2017).
\newblock Proceedings of the 1st International Conference on the Material Point
  Method (MPM 2017)

\bibitem{Toro}
Toro, E.F.: Riemann Solvers and Numerical Methods for Fluid Dynamics.
\newblock Springer (2009)

\bibitem{Tran}
Tran, L.T., Kim, J., Berzins, M.: Solving time-dependent pdes using the
  material point method, a case study from gas dynamics.
\newblock International Journal for Numerical Methods in Fluids \textbf{62}(7),
  709--732 (2010)

\bibitem{LargeDeform2}
Tran, Q.A., Sołowski, W.: Generalized interpolation material point method
  modelling of large deformation problems including strain-rate effects –
  application to penetration and progressive failure problems.
\newblock Computers and Geotechnics \textbf{106}, 249--265 (2019)

\bibitem{mu_art_1}
Von~Neumann, J., Richtmyer, R.: A method for the numerical calculation of
  hydrodynamic shocks.
\newblock Journal of Applied Physics \textbf{21}(3), 232--237 (1950).
\newblock \doi{10.1063/1.1699639}

\bibitem{fratture}
Wang, S., Ding, M., Gast, T.F., Zhu, L., Gagniere, S., Jiang, C., Teran, J.M.:
  Simulation and visualization of ductile fracture with the material point
  method.
\newblock Proc. ACM Comput. Graph. Interact. Tech. \textbf{2}(2) (2019)

\bibitem{MultiGPU}
Wang, X., Qiu, Y., Slattery, S.R., Fang, Y., Li, M., Zhu, S.C., Zhu, Y., Tang,
  M., Manocha, D., Jiang, C.: A massively parallel and scalable multi-gpu
  material point method.
\newblock ACM Trans. Graph. \textbf{39}(4) (2020)

\bibitem{Wobbes}
Wobbes, E., Möller, M., Galavi, V., Vuik, C.: Conservative taylor least
  squares reconstruction with application to material point methods.
\newblock International Journal for Numerical Methods in Engineering
  \textbf{117}(3), 271--290 (2019)

\bibitem{MPMBook}
X.~Zhang Z.~Chen, Y.L.: The Material Point Method.
\newblock Academic Press, Elsevier (2017)

\bibitem{YorkThesis}
York, A.: Development of modifications to the material point method for the
  simulation of thin membranes, compressible fluids, and their interactions
  (1997)

\bibitem{York}
York, A.R., Sulsky, D., Schreyer, H.L.: Fluid–membrane interaction based on
  the material point method.
\newblock International Journal for Numerical Methods in Engineering
  \textbf{48}, 901--924 (2000)

\end{thebibliography}

\end{document}